\documentclass[
 aps,
 11pt,
 final,
 notitlepage,
 oneside,
 twocolumn,
 nobibnotes,
 nofootinbib,
 superscriptaddress,
 noshowpacs,
 centertags]
{revtex4}

\def\saoname{Special Astrophysical Observatory,  Russian Academy of Sciences,
              Nizhnii Arkhyz, 369167 Russia}

\def\squareforqed{\hbox{\rlap{$\sqcap$}$\sqcup$}}

\def\sq{\ifmmode\squareforqed\else{\unskip\nobreak\hfil
\penalty50\hskip1em\null\nobreak\hfil\squareforqed
\parfillskip=0pt\finalhyphendemerits=0\endgraf}\fi}

\def\arcmin{\hbox{$^\prime$}}

\def\arcsec{\hbox{$^{\prime\prime}$}}

\def\utw{\smash{\rlap{\lower5pt\hbox{$\sim$}}}}

\def\udtw{\smash{\rlap{\lower6pt\hbox{$\approx$}}}}

\def\diameter{{\ifmmode\mathchoice
{\ooalign{\hfil\hbox{$\displaystyle/$}\hfil\crcr
{\hbox{$\displaystyle\mathchar"20D$}}}}
{\ooalign{\hfil\hbox{$\textstyle/$}\hfil\crcr
{\hbox{$\textstyle\mathchar"20D$}}}}
{\ooalign{\hfil\hbox{$\scriptstyle/$}\hfil\crcr
{\hbox{$\scriptstyle\mathchar"20D$}}}}
{\ooalign{\hfil\hbox{$\scriptscriptstyle/$}\hfil\crcr
{\hbox{$\scriptscriptstyle\mathchar"20D$}}}}
\else{\ooalign{\hfil/\hfil\crcr\mathhexbox20D}}%
\fi}}

\newcommand{\ab}{Astrophysical Bulletin }

\newcommand{\aap}{Astron. and Astrophys. }

\newcommand{\aj}{Astron.~J. }
\newcommand{\apjs}{Astrophys.~J. Suppl. }

\newcommand{\mnras}{Monthly Notices Royal Astron. Soc. }
\newcommand{\pasa}{Publ. Astron. Soc. Australia }

\newcommand{\pasp}{Publ. Astron. Soc. Pacific }
\newcommand{\arep}{Astronomy Reports }

\begin{document}
\selectlanguage{english}


%

\title{ANALYSIS OF INTEGRATED-LIGHT SPECTRA OF GALACTIC GLOBULAR CLUSTERS}

\author{\firstname{M.~E.}~\surname{Sharina}}
 \email{sme@sao.ru}
 \affiliation{\saoname}

\author{\firstname{V.~V.}~\surname{Shimansky}}
 \affiliation{Kazan Federal University, Kazan, 420008, Russia}

\author{\firstname{N.~N.}~\surname{Shimanskaya}}
 \affiliation{Kazan Federal University, Kazan, 420008, Russia}
 
 \begin{abstract}
We present the results of determination of the age, helium mass
fraction ($Y$) in terms of the used stellar evolutionary models, 
metallicity ([Fe/H]), and abundances of the
elements C, O, Na, Mg, Ca, Ti, Cr and Mn for 26 globular clusters
of the Galaxy. In this work, we apply a method
developed by us that employs medium-resolution integrated-light
spectra of globular clusters and models of stellar atmospheres and
it is supplemented in this paper by the automatic calculation of
microturbulence velocities of stars in the studied objects. Based
on the data obtained for 26 objects, as well as on the results of our
previous studies, it is shown that the abundances of chemical
elements, that we measured, with the exception of carbon, are
consistent with the literature estimates from the analysis of
\mbox{integrated-light} spectra of clusters and from
high-resolution spectroscopic observations of their brightest
stars. Our estimates of [C/Fe] are consistent with the literature
values obtained from the integrated-light spectra of clusters. We
interpret the systematic difference between the derived [C/Fe] for
globular clusters and the literature  
[C/Fe] values for the
brightest stars of the clusters as a change of the chemical
composition in the atmospheres of stars during their evolution.
The estimated absolute ages and average $Y$ for the clusters are in a
reasonable agreement with the literature data from the analysis of
color--magnitude diagrams of the objects.
\end{abstract}

\maketitle

\section{INTRODUCTION}
\label{intro}
The properties of stellar populations of globular
clusters have always served as the basis for stellar evolutionary
theories~(Carney, 2001; Kruijssen et al., 2019). However, despite the
progress in observational and theoretical astrophysics,
the determination of the absolute ages of globular clusters and helium
content in their stars remains a problem (see
e.g.~Charbonnel, 2016). A particular difficulty is associated with the
explanation of the phenomena of multiple stellar populations in
globular clusters~(Gratton et al., 2012). Abundance anomalies of C, N,
O, Na, Al, Mg vary significantly from object to object. The
feature common for many globular clusters is the presence of
anti-correlations in the abundances of light elements \mbox{C--N},
\mbox{Na--O}, \mbox{Mg--Al}. In some massive globular clusters,
stellar populations with different abundances of \mbox
{$r$-process} elements were found. The difficulty in explaining
these phenomena lies in the fact that the changes in the elemental
abundances that occur during the evolution of cluster stars are
superimposed on the anomalies of the primary chemical contents,
which apparently existed at the time of cluster
formation~(Kraft, 1994; Charbonnel, 2016 and references
therein). Therefore, it is important to study integrated-light
spectra of star clusters in order to evaluate how chemical
peculiarities influence the properties of the total radiation of
objects. First of all, it is necessary to study precisely the
globular clusters of our Galaxy, for which there are data from
deep stellar photometry and high-resolution spectroscopy of
individual stars.

\section{OBJECT SAMPLE}
\label{objsel}
Our sample is mainly composed of the globular
clusters with integrated-light spectra from the library of
Schiavon et al. (2005). These authors used the following method of
observations. They drifted the spectrograph slit across the core
diameter of the cluster. The sky background spectra were obtained
in a similar way outside the boundaries of the clusters. The
typical size of the background area was
\mbox{$5\arcmin$--$10\arcmin$}. For the objects NGC\,6205 and
NGC\,7006, we used the archival data obtained with the CARELEC
spectrograph at the \mbox{1.93-m} telescope in the Haute-Provence
Observatory. Instrumental conditions during these observations and
methods of data analysis were described in detail in our previous
papers: Sharina et al. (2013), Khamidullina et al. (2014) and 
Sharina et al. (2018). The seeing was $2\arcsec$--$3\arcsec$ during
the observations of NGC\,6205 and NGC\,7006 (July \mbox{9--10},
2010). The spectra of NGC\,6205 and NGC\,7006 were obtained using
fixed slit positions, i.e. without drifting the spectrograph slit.
For NGC\,6205, the observations were carried out using one slit
position near the cluster center and oriented along the sky
meridian (slit position angle \mbox{$PA = 0$).} The exposure time
was 300~s. The sky background exposure was taken outside the
cluster boundaries with \mbox{$PA = 0$}. For NGC\,7006, one
exposure of 900~s. was taken with \mbox{$PA = 0$}. Since the
cluster has a visible diameter of approximately $2\arcmin$, the
regions of the obtained spectrum above and below the object were
used to subtract the background.

In this paper, ages, helium mass fraction $Y$, metallicity [Fe/H]
and elemental abundances are determined for 26 clusters. It will
be shown in Section~\ref{xi_turb} that the procedure of the automatic 
accounting for microturbulence velocities developed in this study 
does not significantly change the
results obtained by us earlier with fixed values of this parameter. 
Therefore, we combine the results of the studies of the
globular clusters in the Galaxy from our previous articles with
the data obtained in this paper to demonstrate the accuracy of the
aforementioned parameters determination taking the advantage of
a large object sample. We use the previous results for the
following objects: NGC\,2419~(Sharina et al., 2013), NGC\,6229,
NGC\,6779, NGC\,5904~(Khamidullina et al. 2014), NGC\,1904, NGC\,5286,
NGC\,6254, NGC\,6752 and NGC\,7089~(Sharina et al., 2017), NGC\,104,
NGC\,6838, NGC\,6121, NGC\,6341 and NGC\,7078~(Sharina et al. 2018).
In total, we analyze the results for 40 objects in
Section~\ref{comparlit}. 
Note that the chemical abundances were
changed due to the improvement of the method of $\xi_{\rm turb}$
determination for the objects NGC\,1851, NGC\,2298, NGC\,3201,
NGC\,6218, NGC\,6342, NGC\,6522, NGC\,6624 studied in
the above mentioned previous papers. 
We make a comparison of our results 
with the following literature values:
(1) determination
of ages and metallicities of clusters with their color-magnitude
diagrams, (2) determination of metallicities
and chemical abundances by Conroy et al. (2018)
who used integrated-light spectra of Galactic globular
clusters from the library of Schiavon et al. (2005), as
well as 3) high resolution spectroscopic abundances
determination.

\section{Method}
\label{method}
The method of our analysis is described in detail
in the papers:~Khamidullina et al. (2014); Sharina et al.
(2014); Sharina and Shimansky (2020); Sharina et al.
(2013, 2018, 2017). In the paper by~Sharina et al. (2017), spectra
of extragalactic clusters were analyzed with it for the first
time. Let us remind the main details of the method. We analyze
\mbox{integrated-light} spectra of globular clusters using models
of stellar atmospheres for the determination of ages, helium mass
fraction $Y$, metallicity [Fe/H], and chemical abundances in the
studied objects. We calculate synthetic spectra based on the plane
parallel hydrostatic models of stellar
atmospheres~(Castelli and Kurucz 2003). Atomic and molecular spectroscopic
line lists were taken from the website of R.~L.~Kurucz
(\url{http://kurucz.harvard.edu/linelists.html}). The atmospheric
parameters are determined from stellar evolutionary isochrones, and the
calculated synthetic spectra of individual stars are summed
according to a certain mass stellar function. In this work, we
calculate synthetic spectra in the approximation of the local
thermodynamic equilibrium~(LTE) and use the isochrones
by~Bertelli et al. (2008) and stellar mass function by~Chabrier (2005).

A comparison of the shapes and intensities of the observed and
model absorption lines of the Balmer series of hydrogen and the
intensities of the CaI\,4227, and K, and H Ca\,II\,3933.7~\AA, and
3968.5~\AA\ lines\footnote{The line H$\epsilon$ contributes the
H\,Ca\,II line.} makes it possible to determine the isochrone that
best reproduces the observed spectrum. The temperature of Main
sequence~(MS) turnoff stars increases with the decreasing age. At
the same time, the depths of the cores and wings of hydrogen lines
increase. The growth of $Y$ leads to the increase in the mean
luminosity and temperature of horizontal branch~(HB) stars. The
intensities and widths of the hydrogen lines demonstrate the
maximum dependency on the stellar luminosity for the stars of type A0
\mbox{($8700 \le T_{\rm eff} \le 11\,000$~K)}. The further growth of
temperature and luminosity leads to the weakening of the hydrogen
lines due to the intense ionization of hydrogen. This process
develops rather slowly, since the decrease in absorption
coefficients in the Balmer hydrogen lines is partially compensated
by a decrease in opacity in the surrounding Paschen continuum.

When metallicity, age and $Y$ change, the depths of the
nuclei and wings of each of the hydrogen lines H$\delta$,
H$\gamma$, and H$\beta$ change in their own way due to the
differences in the contribution of stars of various luminosities and
spectral classes depending on the wavelength. This fact allows one
to confidently determine the metallicity, age and $Y$. The
Ca\,I/Ca\,II ionization balance, discussed above, depends on the
temperature and density of the medium. Therefore, this balance is
an important additional control indicator that allows one to check
the parameters obtained from the hydrogen lines and refine them if
necessary.

Since atmospheric parameters are based on stellar evolution
isochrones, it is important to verify, for the method evaluation purposes,
whether the stellar evolution models selected by analyzing the
integrated light spectra from clusters correspond to the
observed position of the stars in the color--magnitude
diagrams~(CMD) of these objects. This task has been carried out
for all studied Galactic clusters. We described the methods of CMD analysis
 in the paper by~Khamidullina et al. (2014). Since we
know metallicity, age and $Y$ of a cluster from the analysis of
its spectrum, during the isochrones fitting process, we 
only need to choose the distance to a given object and color excess $E(B-V)$
from Galactic color extinction maps~(Schlegel et al., 1998). In order to
test our method,~Khamidullina et al.~(2014) also solved the problem of
selecting the following five parameters using CMD: metallicity, age, $Y$,
$E(B-V)$ and distance to the cluster. However, in this paper, this
approach is not applied.

To summarize, in this article we present the results of modeling
the integrated light spectra of 26 Galactic globular clusters and
compare the results with literature values. The main goal of this
investigation in the framework of our study is further improvement
of our method. Compared to our previous works, we take a new
approach to calculating $\xi_{\rm turb}$.

\begin{figure*}[]
 \setcaptionmargin{-5mm} \onelinecaptionstrue \captionstyle{normal}
\includegraphics[scale=0.9]{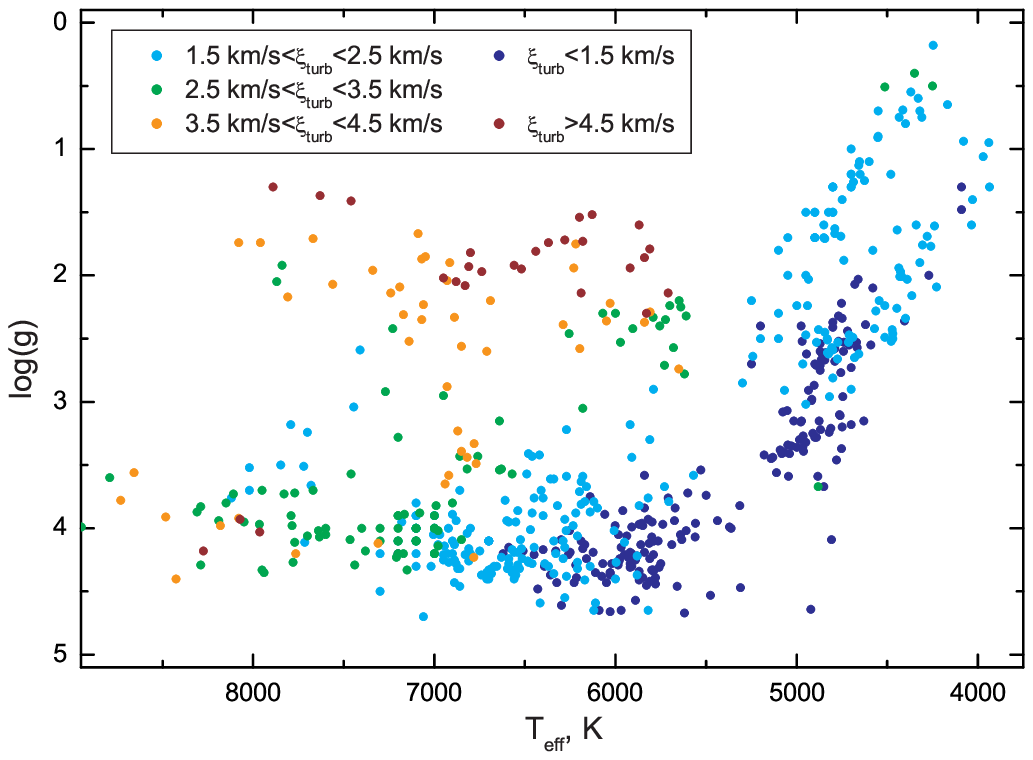} 
\includegraphics[scale=0.9]{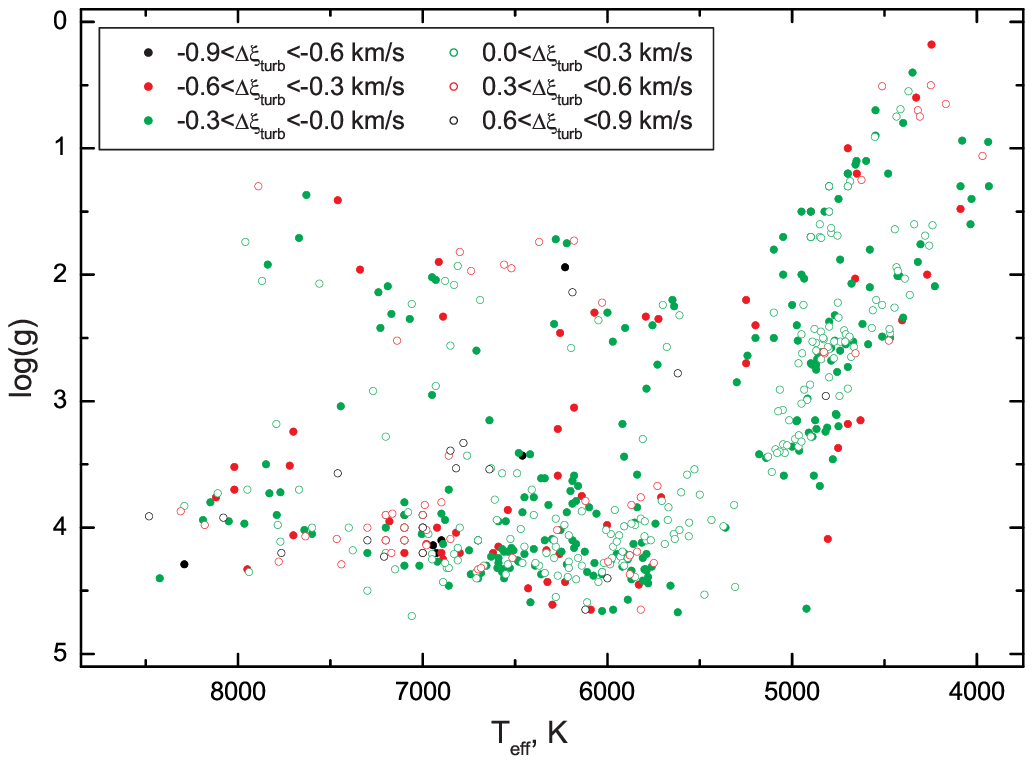} 
 \caption{Panel~(a): the distribution of 637 stars of the sample in the plane $ T_{\rm eff}$--$\log g$. 
 The intervals of $\xi_{\rm turb}$ are indicated where the objects lie.
 Panel~(b): the distribution of 607 stars left after the selection procedure in the plane $T_{\rm eff}$--$\log g$
 (see Section~\ref{xi_turb}). These stars were used to build the function~\ref{microt_eq} (see also Table~\ref{coeff}). 
 The deviation intervals of the calculated $\xi_{\rm turb}$ from the corresponding literature 
 values are indicated.}
 \label{figXiTurb}
\end{figure*}

\begin{figure}[]
 \setcaptionmargin{5mm} \onelinecaptionstrue \captionstyle{normal}
 \vspace{-0.5cm}
 \includegraphics[scale=0.27]{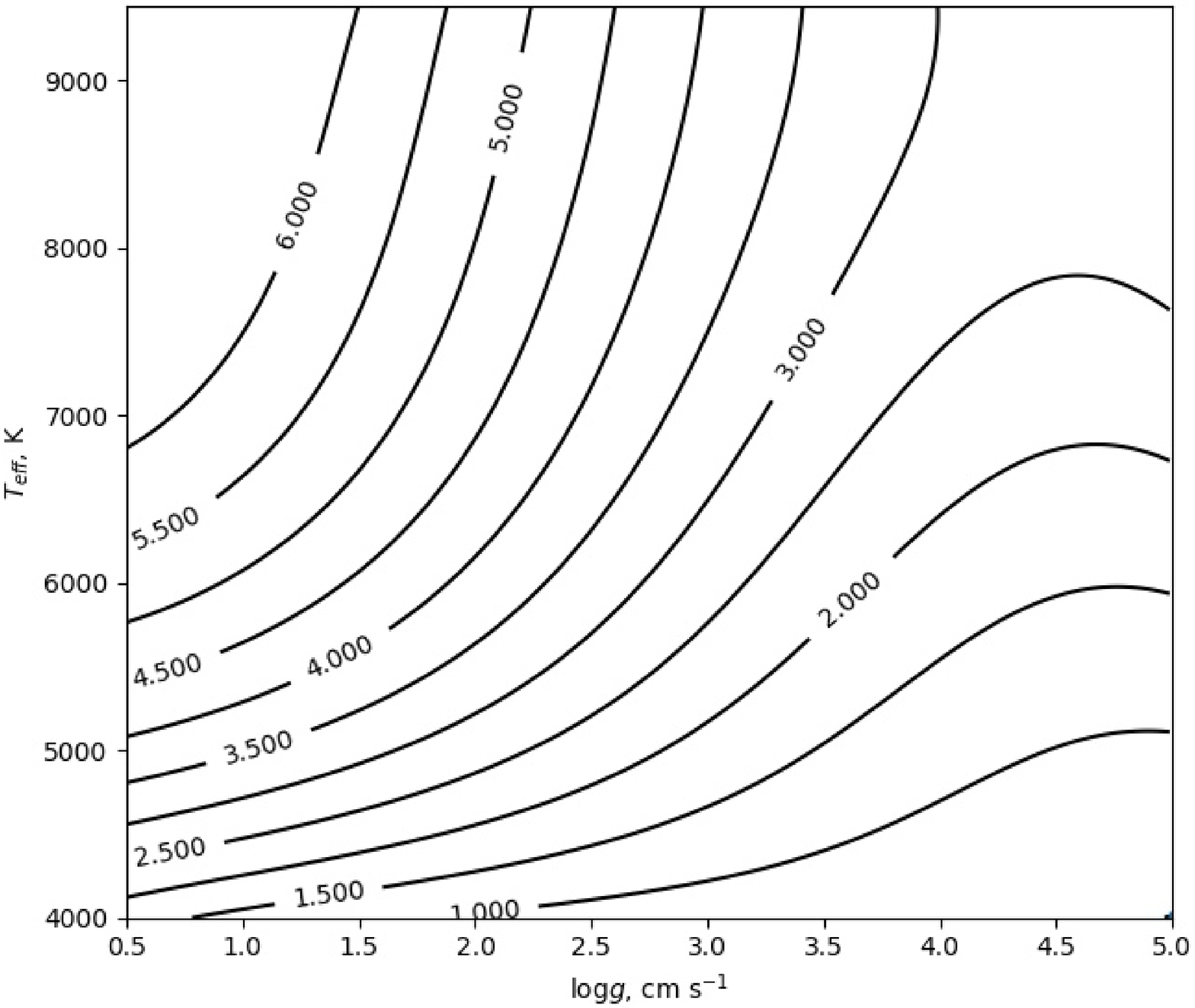}
 \includegraphics[scale=0.33]{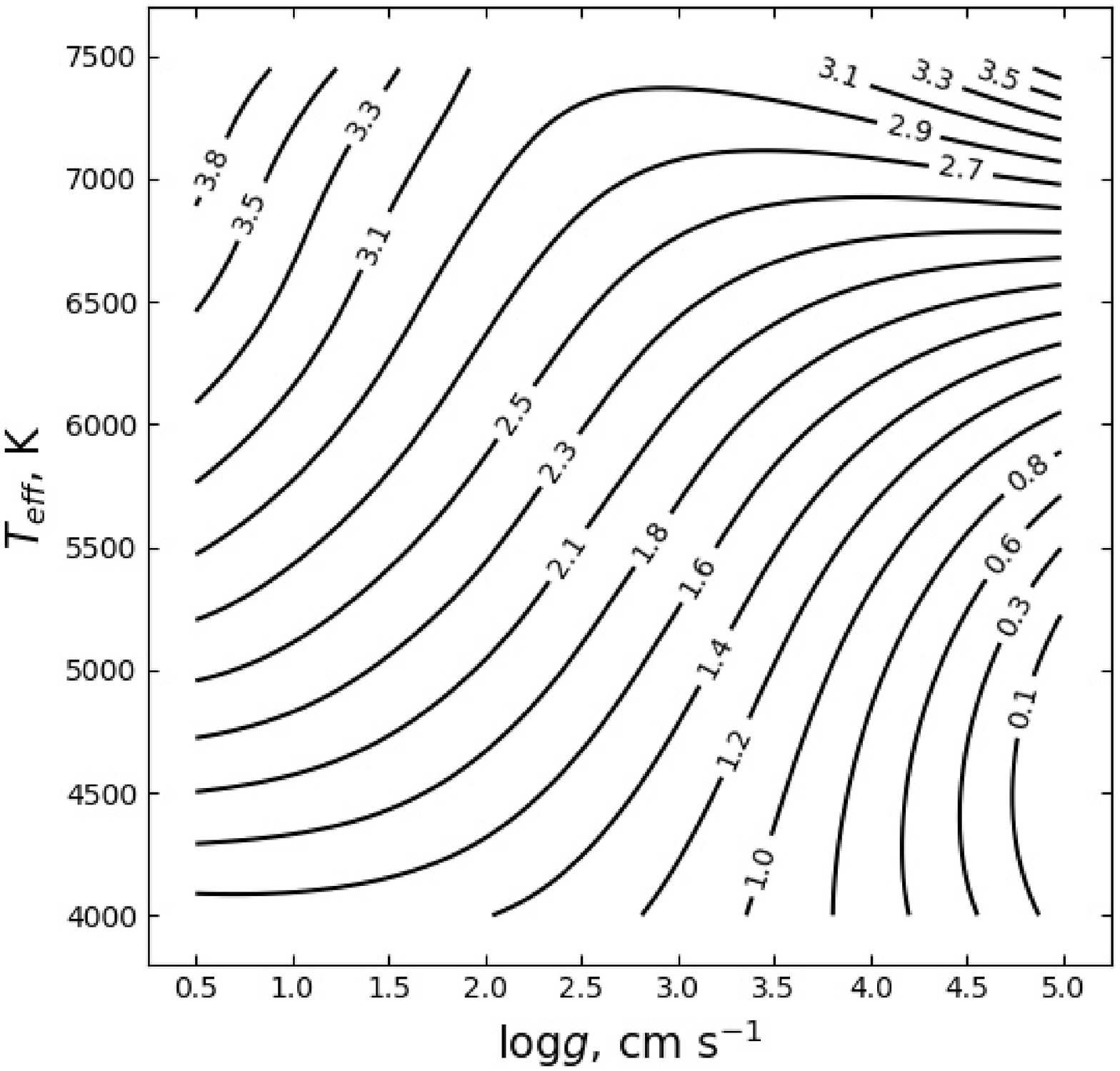}
 \includegraphics[scale=0.33]{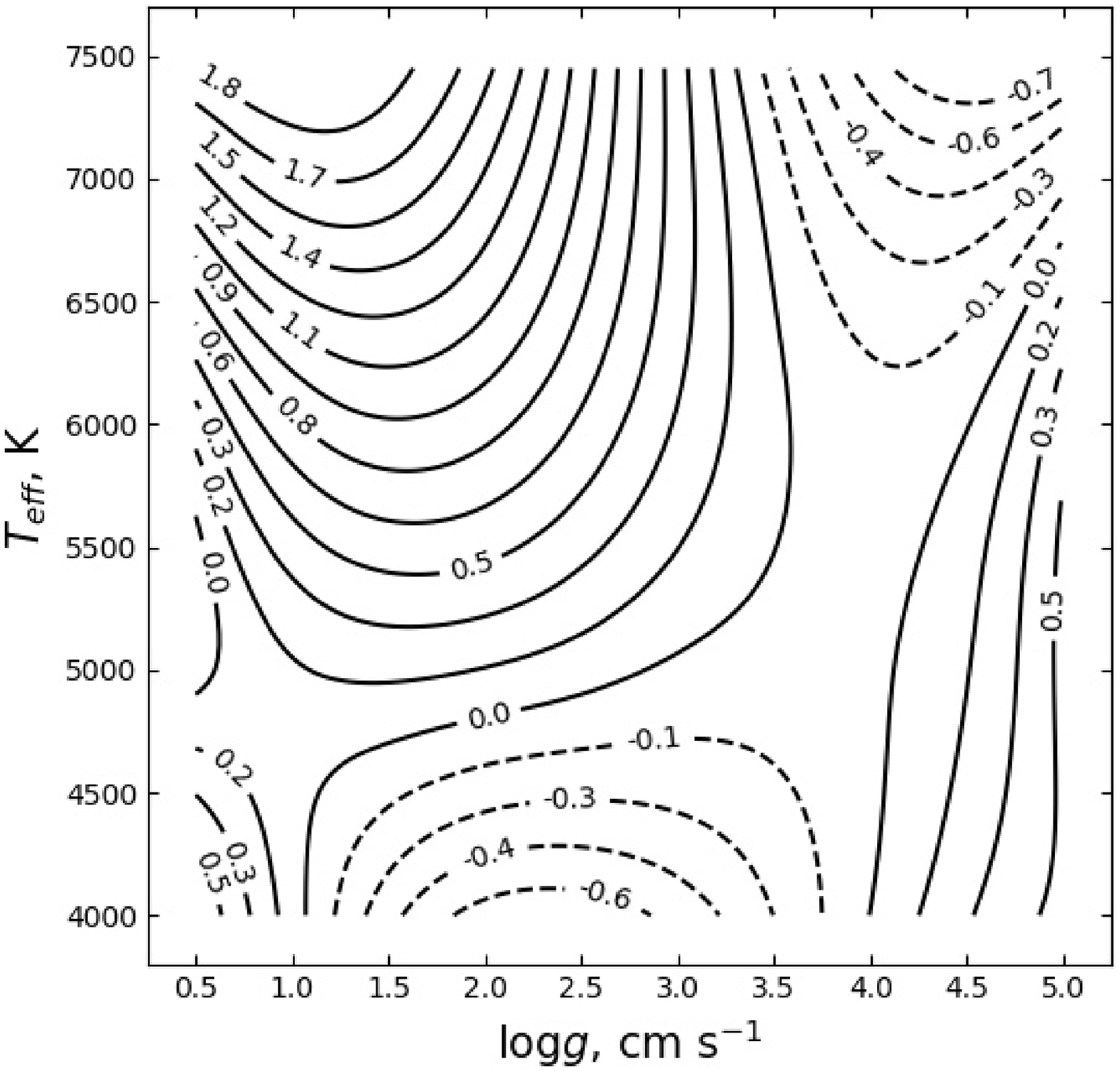}
 \caption{Comparison of the distributions of 
$\xi_{\rm turb}$ values obtained by us and Boeche and Grebel (2016) in
contours on the plane \mbox{$ T_{\rm eff}$--$\log g$}. Panel~(a):
the distribution of the $\xi_{\rm turb}$ values obtained using
the function~(\ref{microt_eq}). Panel~(b): the distribution of
the Boeche and Grebel~(2016) $\xi_{\rm turb}$ values. Panel~(c): the
difference in km\,s$^{-1}$ ($\Delta \xi_{\rm turb}$) between
$\xi_{\rm turb}$ values obtained by us and Boeche and Grebel~(2016). Contours with
negative $\Delta \xi_{\rm turb}$ are shown in dashed lines.}
 \label{figXiTurb1}
\end{figure}

\subsection{Microturbulence velocity}
\label{xi_turb}

In our previous studies of integrated-light spectra of globular
clusters (Khamidullina et al., 2014; Sharina et al., 2014, 2013, 2018, 2017), 
the same microturbulence velocity value was set for all
cluster stars based on the best agreement between theoretical and
observational data. However, this approach is a priori incorrect,
because $ \xi_{\rm turb}$ values vary in a wide range for cluster
stars with different stellar parameters. Therefore, we have
developed and implemented a method for automatic selection of
$ \xi_{\rm turb}$ values when calculating synthetic spectra of
stars. The aim is to take into account their real variations in
clusters and reduce the number of free parameters in the
analysis of integrated-light spectra. Within the framework of this
technique, the functional dependence of $ \xi_{\rm turb}$ on the following main
parameters of stellar atmospheres is built: the effective
temperature $ T_{\rm eff}$ and the gravity on the stellar surface
$ \log g$. For the function building, the observational parameters
of stars from the literature were used. In the analyzed sample for
parameter ranges, we included 637 objects from complex
spectroscopic studies of stars~(Bruntt et al., 2012; Kahraman Alicavųs
et al., 2016; Menzhevitski et al., 2014; Santos
et al., 2013; Schaeuble et al., 2015) where microturbulence velocities were
derived from the analysis of Fe\,I and Fe\,II lines using the model
atmospheres method and satisfying the standard
condition that there should be no dependence between the iron abundance
on the equivalent widths of individual lines. The $ \xi_{\rm
turb}$ values obtained in the literature using different methods
did not show significant differences for stars of the same
spectral class. Furthermore, it turned out that 
the functional dependence of $ \xi_{\rm turb}$ on $ T_{\rm eff}$ and $\log g$ 
is smooth during the transition from one spectral
class to another, which allowed describing it by a third-degree polynomial:

\begin{equation}\label{microt_eq}
\xi_{\rm poly}=\sum_{i,j=0}^3 a_{ij} \left(\dfrac{T_{\rm
eff}}{1000~K}\right)^i(\log g)^j,
\end{equation}
where the coefficients calculated using the least squares method
are presented in Table~\ref{coeff}. A similar relationship was
built by~Boeche et al. (2011), but for a narrower range of the
parameters $T_{\rm eff}$ and $\xi_{\rm turb}$. The approximation
of the observed data in~Boeche et al. (2011) was performed by 
a second-degree polynomial function taking into account all the cross
components. Earlier,~Boeche et al. (2011) constructed the function
of $ \xi_{\rm turb}$ versus $ T_{\rm eff}$ and $\log g$ using a
third-degree polynomial. To our knowledge, except for the studies
of~Boeche et al. (2011) and ours, there are no there were no other attempts to construct
a three-dimensional dependence for stars in a wide
range of $ T_{\rm eff}$ and $\log g$ (Boeche et al., 2011; Larsen et al., 2012; Malavolta
et al., 2014; Marino et al., 2008; McWilliam
and Bernstein, 2008; Nissen, 1981 and references in this
papers.)

\begin{table*}[]
\setcaptionmargin{0mm} \onelinecaptionstrue \captionstyle{normal}
\caption{Coefficients of the third-degree polynomial function~(\ref{microt_eq}).} \label{coeff}
\medskip
\begin{tabular}{c|c|c|c|c}
\hline
         &  $a_{i0}$     &  $a_{i1}$     &  $a_{i2}$     &   $a_{i3}$ \\
\hline
$a_{0j}$ & $0.943161$          & $0.296562$          & $2.47216{\rm E}-2$  & $-1.69846{\rm E}-3$ \\
$a_{1j}$ & $-0.154776$         & $-0.128602$         & $1.38775{\rm E}-2$  & $-5.05178{\rm E}-4$  \\
$a_{2j}$ & $0.106128$          & $-3.58510{\rm E}-2$ & $4.91499{\rm E}-3$  & $-1.43992{\rm E}-4$  \\
$a_{3j}$ & $-9.99955{\rm E}-3$ & $5.57946{\rm E}-3$  & $-9.46429{\rm E}-4$ & $3.54326{\rm E}-5$  \\
  \hline
\end{tabular}
\end{table*}

To obtain a sustainable solution, we set the following boundary
conditions: $\xi_{\rm turb}=0.55$~km\,s$^{-1}$ for $T_{\rm
eff}=4000$~K, $\log g=5.0$ and $\xi_{\rm turb}=7.0$~km\,s$^{-1}$
for $T_{\rm eff}=10\,000$~K, $\log g=0.5-1.0$~(Galazutdinov et al. 2017)
by adding artificial points with the appropriate values $ T_{\rm
eff}$, $\log g$ and $\xi_{\rm turb}$ and specially selected
weights. The set of observed values was approximated by a
two-dimensional polynomial where the coefficients for all
degrees of the components with respect to $ T_{\rm eff}$ and $\log
g$ (including the cross components) were calculated using the
least squares method. The global minimum of the sum of the squared
deviations of the weighted observed $\xi_{\rm turb}$ and
aproximated $\xi_ {\rm poly}$ values of the microturbulence
velocity was found by differentiating it with respect to all
coefficients of the polynomial function and forming for them a connected
coherent system of linear equations
\begin{equation}\label{microt_eq1} \dfrac{d(w(\xi_{\rm turb}-\xi_{\rm poly})^2)}{d(a_{ij})} = 0~~~~~~~i,j=0-3, \end{equation}
where $w$---weight, $\xi_{\rm turb}$---microturbulence velocity
from the literature, $\xi_{\rm poly}$---microturbulence velocity
defined by the polynomial fit, $ a_{ij}$---coefficients of the polynomial (see
Table~\ref{coeff}). In the process of approximation, iterative
rejection of the observed values deviating from the polynomial function by
more than $3\sigma$ of the current dispersion was carried out. In
addition, each point was assigned a weight $w$ equal to $\dfrac{1}
{\sqrt n_{\rm loc}}$, where $n_{\rm loc}$ was determined by the
number of points in sectors with \mbox{$\Delta T_{\rm eff} =
500$}~K  and \mbox{$\Delta \log g = 0.5$}. As a result, a uniform
smooth distribution of $\xi_{\rm turb}$ was obtained for the 607
stars for the ranges \mbox{$T_{\rm eff} = 4000$--$10\,000$~K} and
\mbox{$\log g = 0.5$--$5.0$} with a standard error of
0.27~km\,s$^{-1}$ and the absence of dependence of this error on the
atmospheric parameters.

The location of 637 sample stars on the plane \mbox{$T_{\rm
eff}$--$\log g$} is shown in Fig.~\ref{figXiTurb}a with the
intervals $\xi_{\rm turb} $ where the objects lie.
Fig.~\ref{figXiTurb}b shows the distribution on the plane $T_{\rm
eff}$--$\log g$ of the sample stars \mbox{($N=607$)} used to
construct the functional dependence~(\ref{microt_eq}) with the deviation
intervals of the calculated $\xi_{\rm turb}$ values from the
corresponding literature observational values: \mbox{$\Delta
\xi_{\rm turb} =  \xi_{\rm turb}^{\rm our} - \xi_{\rm turb}^{\rm
obs}$}. We have taken into account a relatively small number of
cold \mbox{$T_{\rm eff} <5500$~K} MS stars. However, the available
variations of $\xi_{\rm turb}$ in the atmospheres of such stars
(no more than \mbox{$\Delta \xi_ {\rm turb}=0.3$~km\,s$^{-1})$}
and the presence of a rigidly fixed value at \mbox{$T_{\rm eff} =
4000$}~K, \mbox{$\log g= 5.0$} allows one to calculate the
microturbulence velocity with an average error of less than
0.27~km\,s$^{-1}$. In particular, for the solar atmosphere, the
built function predicts the value \mbox{$\xi_ {\rm
turb}=1.12$}~km\,s$^{-1}$, which is 0.22~km\,s$^{-1}$ more than
the generally accepted value. The maximum density of the points we
take into account is observed in the region of MS stars with
5500~K $<T_ {\rm eff}<7000$~K, red subgiants and giants, which
provides an average approximation error for them of about
0.22~km\,s$^{-1}$. The density of the points is much lower in the regions of
stars with higher temperatures and luminosities, and
the observed $\xi_{\rm turb}$ values in them often vary up to
\mbox{$\Delta \xi_{\rm turb} = 0.8$}~km\,s$^{-1}$ even at close
atmospheric parameters. Therefore, the standard deviation of the
calculated $\xi_{\rm turb}$ from the corresponding observed values
in these regions exceeds 0.3~km\,s$^{-1}$. The largest deviation
amplitude of \mbox{$\Delta \xi_ {\rm turb} = 0.39$}~km\,s$^{-1}$
is observed in the MS region with temperatures of \mbox{$6900
<T_{\rm eff}<7600$}~K and may be due to the inclusion of several
Am-stars with low turbulence velocities. In general, we can
conclude that the obtained distribution allows us to determine the
best $\xi_{\rm turb}$ values for groups of stars that make the
main contribution to integrated spectra of globular clusters (F
and G dwarfs of the MS, subgiants, and red giants).

Figure~\ref{figXiTurb1}a shows in contours the distribution of
$\xi_{\rm turb}$ values obtained using the function~\ref{microt_eq} 
(see also Table~\ref{coeff}) 
for the entire range of parameters $T_{\rm eff}$ and
$\log g$. It is characterized by the presence of the expected
systematic increase in the microturbulence velocity with the
increase in the effective temperature and luminosity of stars. The
low degree approximating polynomial allowed us to exclude
\mbox{small-scale} fluctuations of the $\xi_{\rm turb}$ values
observed in Fig.~\ref{figXiTurb} in the region of hot stars that are
probably due to methodological differences in determination of
$\xi_{\rm turb}$ in different works.

Figure~\ref{figXiTurb1}b shows in contours the distribution of
$\xi_{\rm turb}$ values on the plane \mbox{$T_{\rm eff}$--$\log
g$} according to the function from the paper by~Boeche and Grebel (2016).
Figure~\ref{figXiTurb1}c shows the difference between our
distribution of $\xi_{\rm turb}$ values and that obtained
in~Boeche and Grebel~(2016).

The distribution of $\xi_{\rm turb}$ values in the~Boeche and Grebel~(2016)
sample has the following important differences from
our distribution: 1)~the stars are mostly concentrated on the MS
within the temperature range \mbox{$5000<T_{\rm eff}<6500$~K} and
on the sub-giant and red giant branches with \mbox{$1.5 < \log g <
3.5$}; 2)~there are ten stars with the temperatures \mbox{$6500 <
T_{\rm eff} <7000$}~K ; 3)~the region of \mbox{F-giants} and super
giants contains 20 objects with the parameters in the range
\mbox{$5600<T_{\rm eff}<6600$}~K, \mbox{$1.0 < \log g < 2.6$};
4)~there are no data between the indicated areas, as well as for the
temperatures higher than \mbox{$T_{\rm eff}=7000$}~K.

Note that the function built
in~Boeche and Grebel~(2016) yields an increase in microturbulence
velocity with increasing temperature and luminosity of stars,
close to that found in our work. Our function and that of
Boeche and Grebel~(2016) yield the best
agreement in the area of the highest data density
in~Boeche and Grebel~(2016): on the cold side of the MS and on the
branches of red subgiants and giants where the differences in the
$\xi_{\rm turb}$ values do not exceed 0.3~km\,s$^{-1}$. Note that
in these areas our function yields on average lower (by
\mbox{0.05--0.15}~km\,s$^{-1}$) microturbulence velocity values
than the function of~Boeche and Grebel~(2016). Outside this region, the
differences between the two distributions rapidly increase to
0.4~km\,s$^{-1}$ for the brightest red giants, up to
0.5~km\,s$^{-1}$ for MS stars with temperatures \mbox{$ T_{\rm
eff}>7000$}~K, and up to 1.9~km\,s$^{-1}$ for F supergiants. These
differences are due to the limited set of $\xi_{\rm turb}$ values
in~Boeche and Grebel~(2016), which are either small in number or
completely absent in these areas. For MS stars with temperatures
\mbox{$T_ {\rm eff}<5000$}~K, the function by~Boeche and Grebel~(2016)
can yield $\xi_{\rm turb}$ values that are negative or close to zero due to
the absence of a fixed boundary value. As a result, we can
conclude that our functional dependence of the microturbulence velocity
on atmospheric parameters provides correct estimates of
$\xi_{\rm turb}$ in wide ranges of $T_{\rm eff}$ and $\log g$. In
the areas of F- and G-dwarfs, red subgiants and giants, the
results of applying our and~Boeche and Grebel~(2016) functions
correspond to each other within the errors of these functions calculation.

Our function~(\ref{microt_eq}) (see also Table 1) is integrated into
the program for the computing of \mbox{integrated-light} spectra of
clusters.  If any of the parameters go out of the range, we consider 
 $\xi_{\rm turb}$  to be equal in its value
to the nearest boundary point. Modeling the spectra of
globular clusters NGC\,6254 and NGC\,6341 using the developed
$\xi_{\rm turb}$ distribution leads to the average values of the
microturbulence velocities: $\xi_{\rm turb} =
1.86$~km\,s$^{-1}$  and \mbox{$\xi_{\rm turb} =
1.88$}~km\,s$^{-1}$, correspondingly. These values are consistent within
0.20~km\,s$^{-1}$ with the corresponding data found for NGC\,6254
and NGC\,6341 in our previous works~(Khamidullina et al., 2014; Sharina et al.,
2018, 2017).

\begin{figure*}[]
 \setcaptionmargin{5mm} \onelinecaptionstrue \captionstyle{normal}
\begin{tabular}{p{0.325\textwidth}p{0.325\textwidth}p{0.325\textwidth}}
 \includegraphics[scale=0.27,angle=-180]{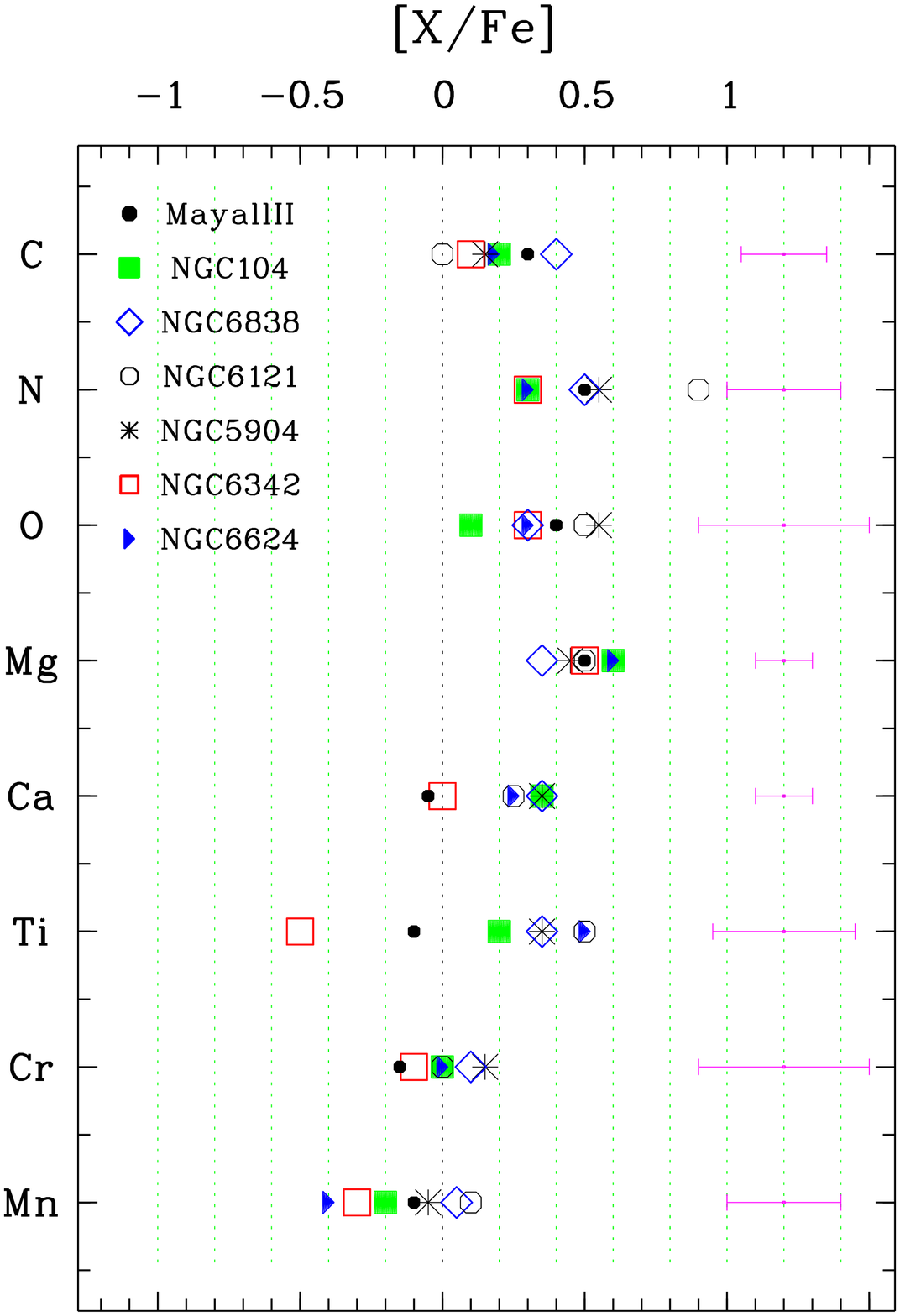} &
\includegraphics[scale=0.27,angle=-180]{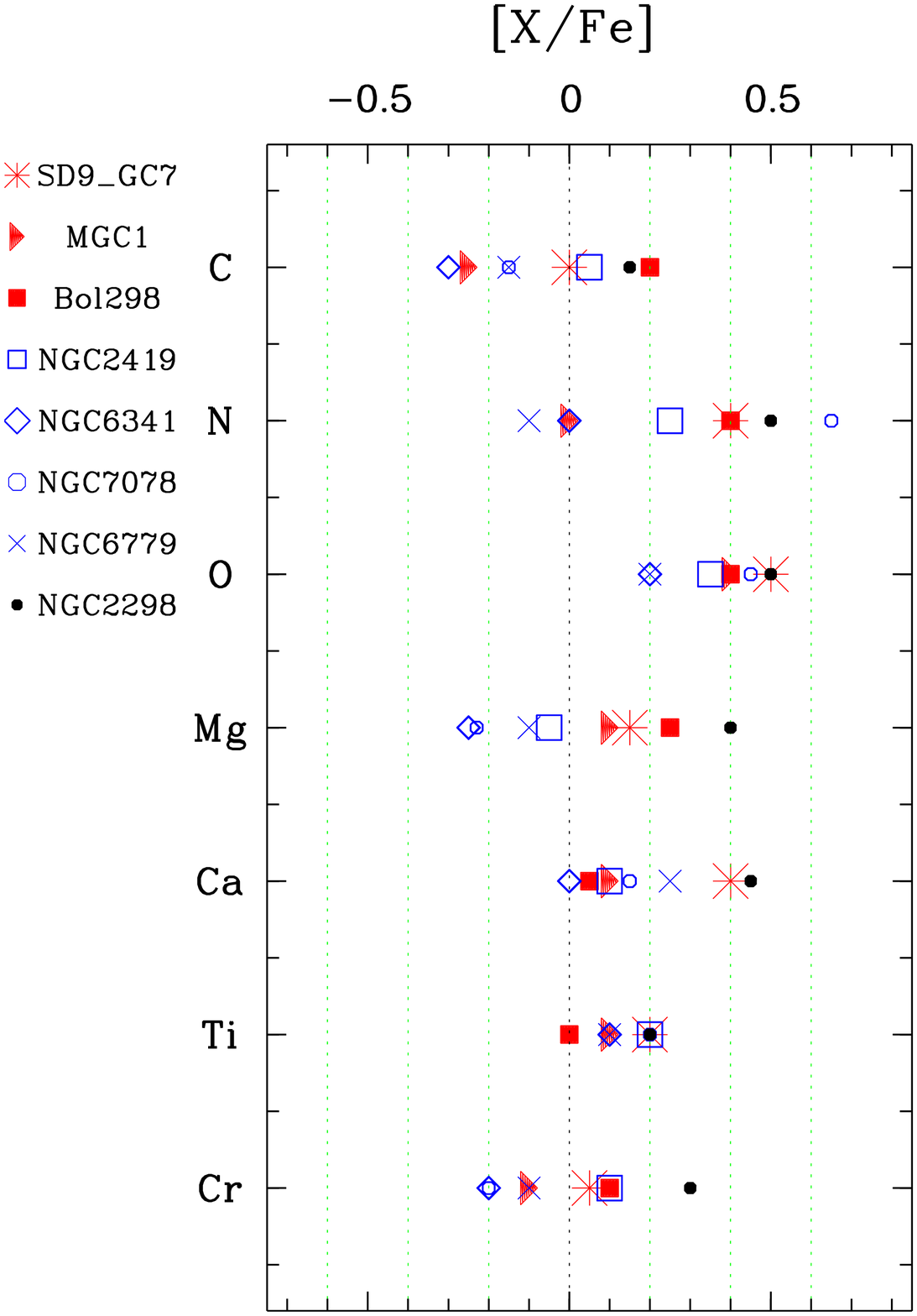} &   %
\includegraphics[scale=0.27,angle=-180]{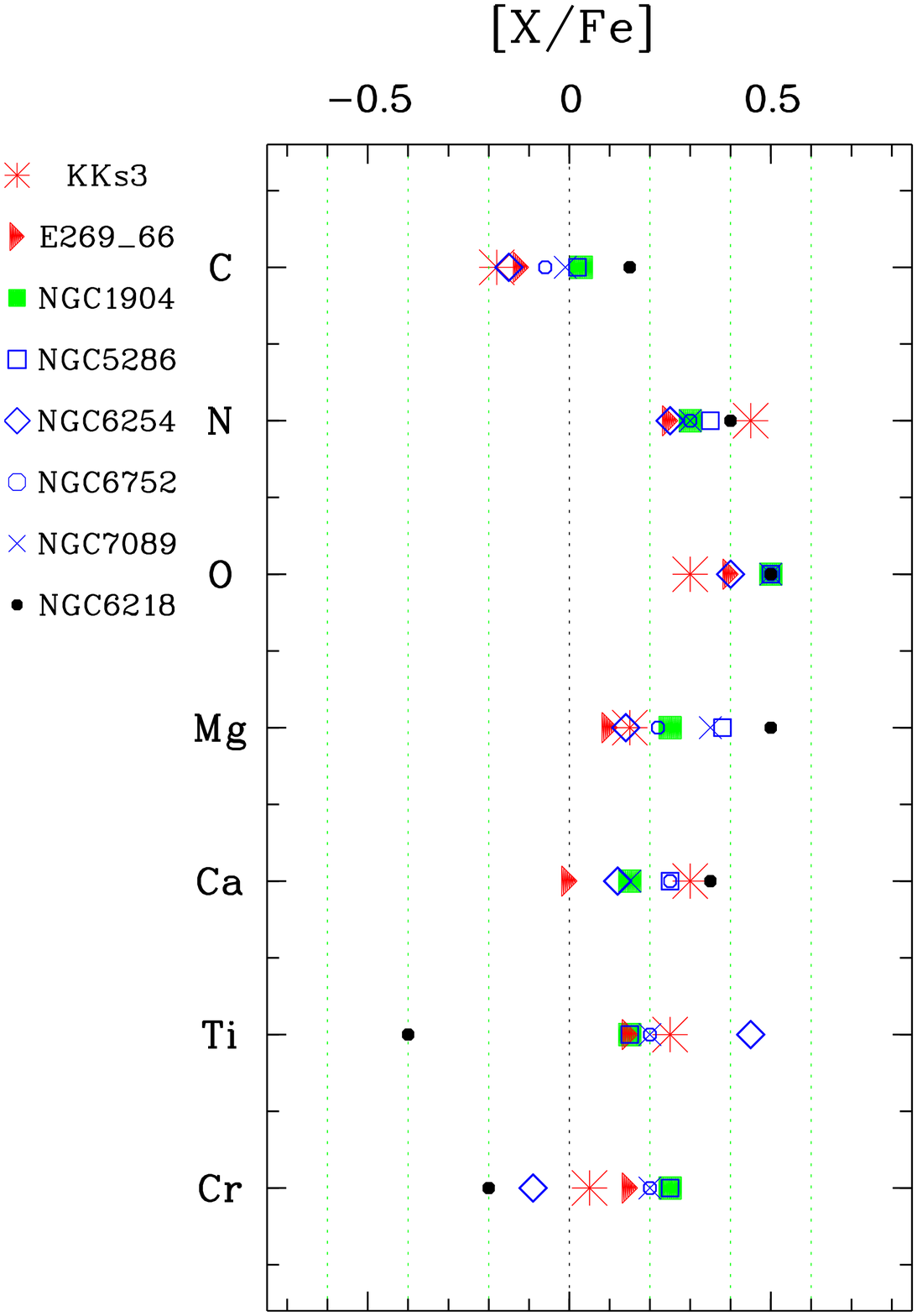} \\
\includegraphics[scale=0.27,angle=-180]{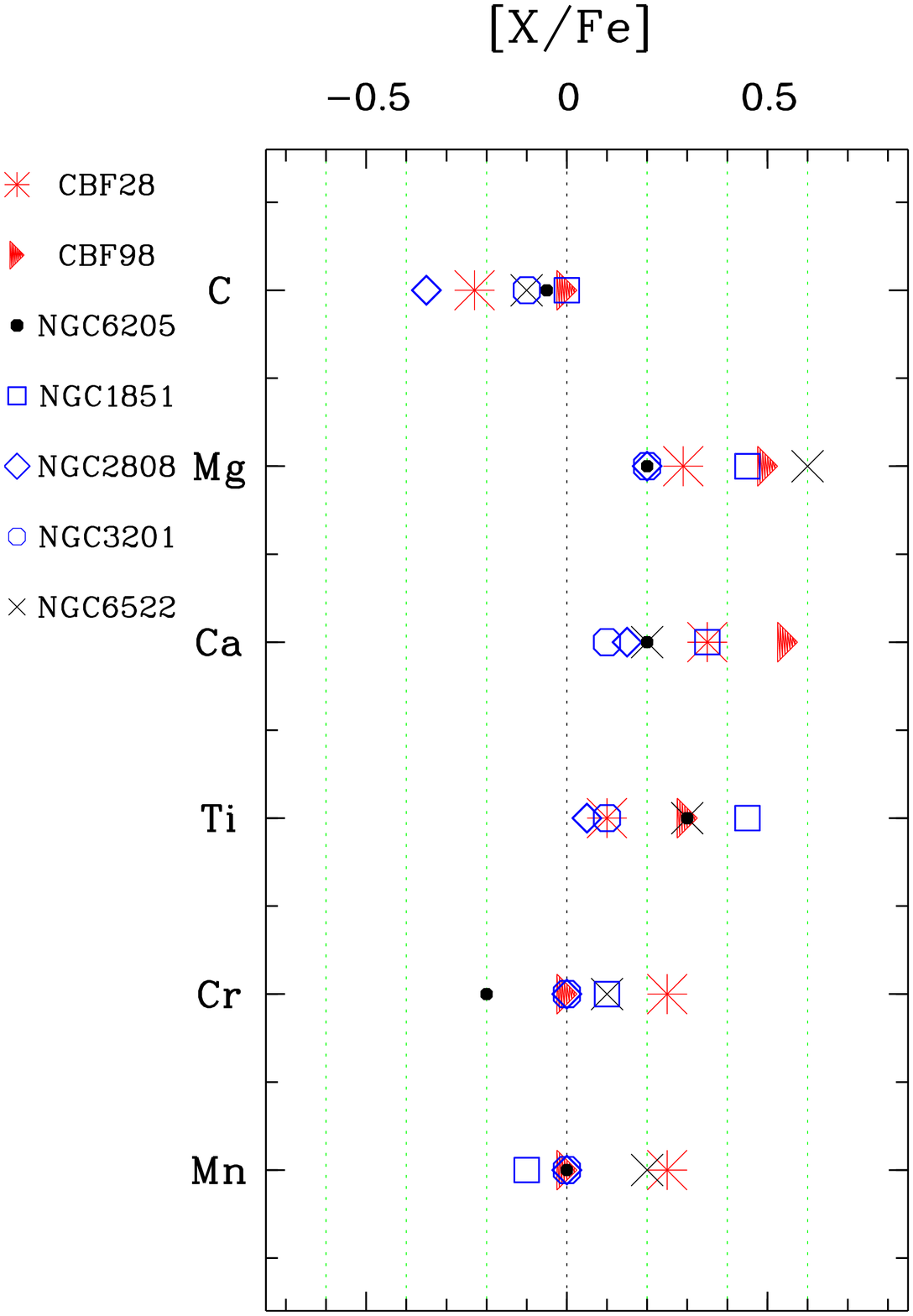} &
 \includegraphics[scale=0.27,angle=-180]{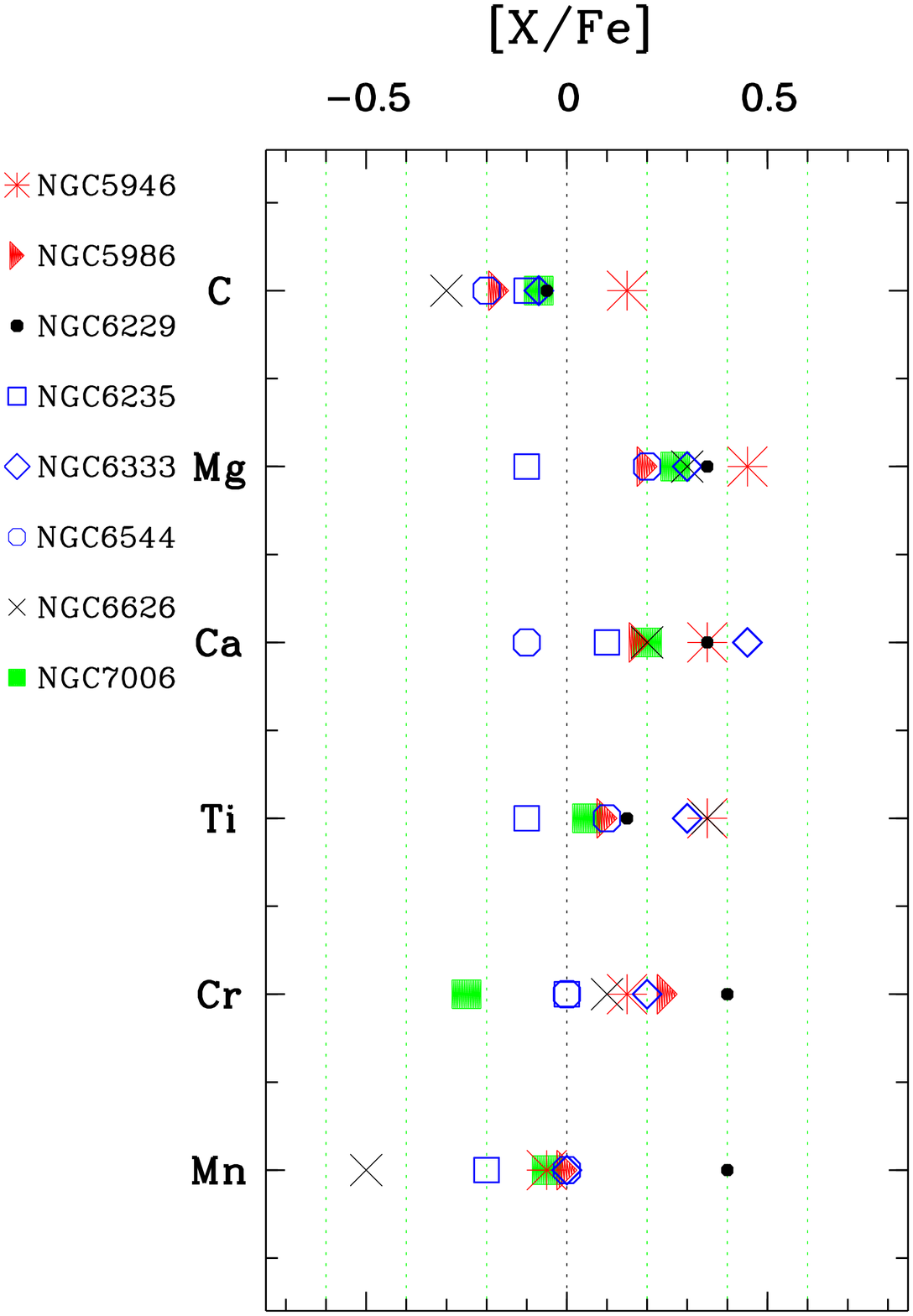} &
 \includegraphics[scale=0.27,angle=-180]{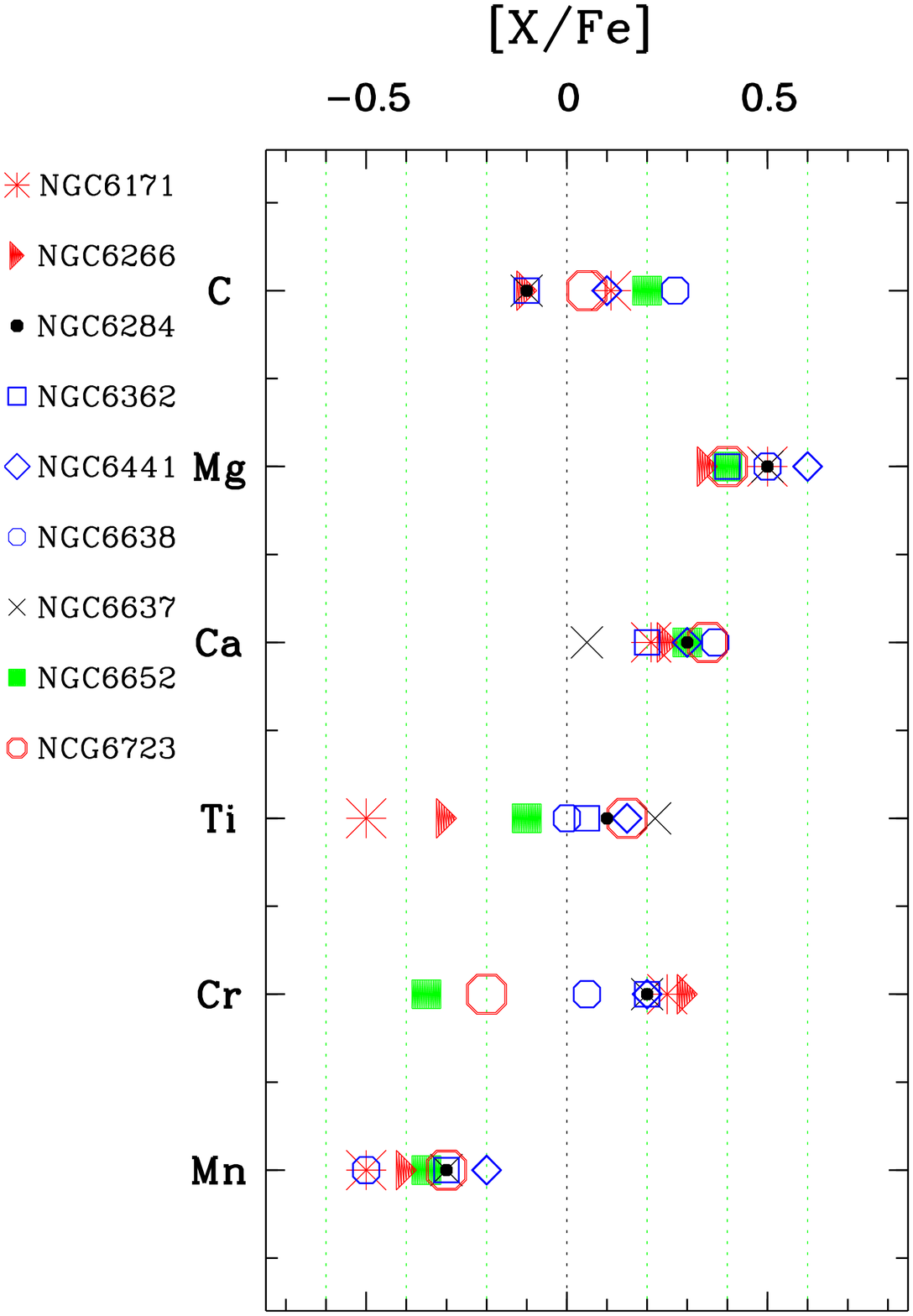} \\
 \end{tabular}
 \caption{Chemical abundances determined using our method 
 based on the determined evolutionary parameters (Table~\ref{tab_abund}). 
The objects are grouped according to [Fe/H]. Typical abundance errors are shown in the panel~(a). }
 \label{fig_Elem}
\end{figure*}

 \begin{figure*}[]
 \setcaptionmargin{-5mm} \onelinecaptionstrue \captionstyle{normal}
\includegraphics[scale=0.7]{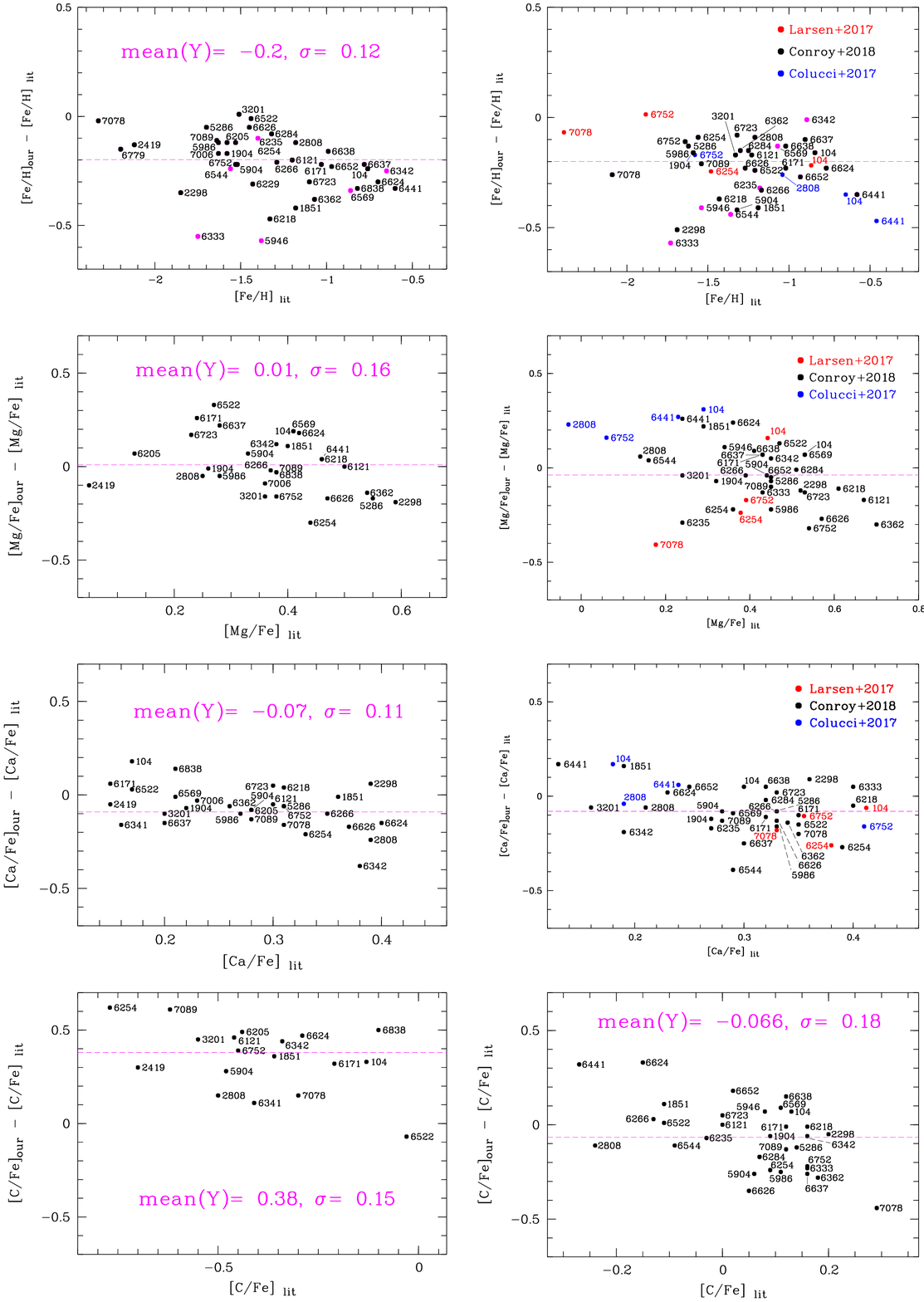} 
 \caption{Comparison of the chemical abundances determined by us for Galactic GCs
 with the high-resolution spectroscopy results from the literature
 (see references in the text) (left panels) and with the literature results obtained
 using integrated-light spectra of the clusters (right panels).
 The objects with low $S/N$ in their spectra are shown as magenta dots.
 \mbox{NGC-numbers} of the clusters are indicated.}
 \label{fig_abund}
\end{figure*}

\begin{table*}[]
\setcaptionmargin{0mm} \onelinecaptionstrue \captionstyle{normal}
\caption{
Properties of Galactic globular clusters estimated using
the analysis of stellar photometry results in this paper and in
the literature: (2)~metallicity from the work
of~Harris~(1996)~(H); (3)~metallicity from the paper
of~VandenBerg et al.(2013)~(V); (4)--(7)~age, helium mass fraction $Y$,
apparent distance modulus, Galactic extinction
(\mbox{color-excess)} according to~VandenBerg et al.(2013) (or in the
absence of them~---from the articles of~Harris~(1996)~(H),
Kruijssen et al. (2019)~(K) and Testa et al.~2001~(T)). In the last three
columns, our results are presented: distance modulus corrected for
the Galactic extinction, \mbox{color-excess} and the parameters of
the evolutionary isochrones~(Bertelli et al., 2008) used for the spectra
modeling: metallicity $Z$, helium mass fraction $Y$ and logarithm
of the age in Gyr. If the cluster spectrum can be adequately
described using two isochrones, the index ``CMD'' indicates the
isochrone that better describes the CMD of the cluster}
\label{photMWGCs}
\scriptsize
\begin{center}
\begin{tabular}{|c|r|c|l|c|l|l|c|c|l|}       
\hline 
Parameter/  & [Fe/H]$^{\rm H}$& [Fe/H]$^{\rm V}$& T$^{\rm V}$    &  Y$^{\rm V}$    & $\rm(m\!-\!M)^{\rm V}$ & E(B-V)$^{\rm V}$& $\rm(m-M)_0^{\rm our}$& E(B-V)$^{\rm our}$& Isochrone      \\
 Object    & (dex)      & (dex)         & (Gyr)    &               & (mag)         & (mag)         & (mag)         & (mag)          &                \\ \hline
  (1)      & (2)        &  (3)          & (4)          & (5)           & (6)           & (7)           & (8)           & (9)            & (10)           \\
\hline                                                                                                                                                  
NGC\,1851  & -1.22      & -1.18         & 11.00        & 0.25          & 15.43         & 0.034         & 15.25         & 0.04           &  0.001,0.26,10.10  \\
 NGC\,2298 & -1.85      &  --           & 12.84 $^{\rm K}$   & --            & --            & --            & 14.9          & 0.23           & 0.0004,0.26,10.05  \\
 NGC\,2808 & -1.15      & -1.18         & 11.00        & 0.25          & 15.53         & 0.227         & 14.6          & 0.2            & 0.0004,0.30,10.15  \\
           & -1.15      & -1.18         & 11.00        & 0.25          & 15.53         & 0.227         &     14.85     &    0.21        &  0.001,0.30,10.15$^{\rm CMD}$\\
 NGC\,3201 & -1.58      & -1.51         & 11.50        & 0.25          & 14.10         & 0.280         & 13.3          &  0.26          & 0.0004,0.30,10.10  \\
 NGC\,5946 & -1.38      &  --           & 11.39 $^{\rm K}$   &  --           & --            & --            & 14.8 -14.9    & 0.13           & 0.0004,0.26,10.15  \\
 NGC\,5986 & -1.58      & -1.63         & 12.50        & 0.25          & --            & --            & 15.45 - 15.4  & 0.305          & 0.0001,0.30,10.05  \\
NGC\,6171  & -1.04      & -1.03         & 12.00        & 0.25          & 14.82         & 0.435         & 14.03         &  0.42          &  0.001,0.26,10.0    \\
NGC\,6218  & -1.48      & -1.33         & 13.00        & 0.25          & 14.05         & 0.225         & 13.75         & 0.205          & 0.0004,0.26,10.10 \\
           & -1.48      & -1.33         & 13.00        & 0.25          & 14.05         & 0.225         & 13.55         &  0.18          &  0.001,0.26,10.15$^{\rm CMD}$ \\
 NGC\,6235 & -1.40      & --            & 11.39 $^{\rm K}$   &  --           & 16.41$^{\rm H}$     & 0.360 $^{\rm H}$      & 15.20         & 0.38           &  0.001,0.30,10.10$^{\rm CMD}$ \\
           & -1.40      & --            & 11.39 $^{\rm K}$   &  --           & 16.41$^{\rm H}$     & 0.360 $^{\rm H}$      & 15.20         &  0.33          & 0.0004,0.26,10.15$^{\rm CMD}$\\
 NGC\,6266 & -1.29      & --            & 11.78 $^{\rm K}$   &  --           & 15.64$^{\rm H}$     & 0.470 $^{\rm H}$      & 14.07         &  0.09          & 0.0004,0.30,10.15  \\
 NGC\,6284 & -1.32      & --            & 11.14 $^{\rm K}$   &  --           & 16.80$^{\rm H}$     & 0.280 $^{\rm H}$      & 15.80         &  0.37          & 0.0004,0.30,10.10  \\
 NGC\,6333 & -1.75      & --            & --           & --            & 15.66$^{\rm H}$     & 0.380 $^{\rm H}$      & --            & --             & 0.0004,0.26,10.10  \\
 NGC\,6342 & -0.65      & --            & 12.03 $^{\rm K}$   & --            &    16.10$^{\rm H}$   &    0.460$^{\rm H}$  & 14.90         & 0.55           &  0.002,0.30,10.0     \\
 NGC\,6362 & -0.95      & -1.07         & 12.50        & 0.25          & 14.58         & 0.076         & 14.46         & 0.08           &  0.001,0.30,10.15   \\ 
 NGC\,6441 & -0.53      & --            & 11.26 $^{\rm K}$   & --            &  16.79 $^{\rm H}$   & 0.470 $^{\rm H}$     & 16.0          & 0.49           &  0.001,0.26,10.0$^{\rm CMD}$\\
           & -0.53      & --            & 11.26 $^{\rm K}$   & --            &  16.79 $^{\rm H}$   & 0.470 $^{\rm H}$     & 15.7          & 0.49           &  0.001,0.30,10.15   \\
 NGC\,6522 & -1.44      & --            & --           & --            & 15.94 $^{\rm H}$    & 0.480 $^{\rm H}$     & 14.57         & 0.10           & 0.0004,0.30,10.10$^{\rm CMD}$ \\
           & -1.44      & --            & --           & --            & 15.94 $^{\rm H}$    & 0.480 $^{\rm H}$     & 14.57         & 0.03           &  0.001,0.26,10.10   \\
 NGC\,6544 & -1.56      & --            & 10.37 $^{\rm K}$   & --            & 14.43 $^{\rm H}$    & 0.730$^{\rm H}$     & 11.90         & 0.17           & 0.0001,0.30,10.15  \\
 NGC\,6569 & -0.86      & --            & --           & --            & 16.85 $^{\rm H}$    & 0.550 $^{\rm H}$     & 15.10         & 0.06           &  0.001,0.23,10.15   \\
           & -0.86      & --            & --           & --            & 16.85 $^{\rm H}$    & 0.550 $^{\rm H}$     & 15.25         & 0.06           &  0.001,0.26,10.05$^{\rm CMD}$   \\
 NGC\,6624 & -0.42      & -0.42         & 11.25        & 0.25          & 15.28         & 0.268         & 14.70         & 0.30           &  0.001,0.23,10.15   \\
 NGC\,6626 & -1.45      & --            & 13-14        & 0.23 $^{\rm T}$     & 14.97 $^{\rm H}$    & 0.400 $^{\rm H}$     & 13.87         & 0.38           &  0.001,0.30,10.10   \\
 NGC\,6637 & -0.70      & -0.59         & 11.00        & 0.25          & 15.23         & 0.163         & 14.80         & 0.18           &  0.002,0.23,10.10  \\
 NGC\,6638 & -0.99      & --            &  --          &  --           & 16.15 $^{\rm H}$    & 0.400 $^{\rm H}$     & 14.7          & 0.07           & 0.001,0.30,10.15   \\
 NGC\,6652 & -0.96      & -0.76         & 11.25        & 0.25          & 15.30 $^{\rm H}$    & 0.090 $^{\rm H}$    & 14.85         & 0.13           &  0.002,0.26,10.15  \\
 NGC\,6723 & -1.12      & -1.1          & 12.50        &  0.25         & 14.73         & 0.070         & 14.65         & 0.09           &  0.001,0.26,10.10   \\
NGC\,6205  & -1.54      & -1.58         & 12.00        & 0.25-0.33     & 14.45         & 0.017         & 14.6          & 0.04           & 0.0004,0.30,10.0    \\
 NGC\,7006 & -1.63      & --            & 12.25 $^{\rm K}$   & --            & 18.24 $^{\rm H}$    & 0.050 $^{\rm H}$     & 17.94         & 0.09           & 0.0004,0.26,10.15   \\
\hline
\end{tabular}
\end{center}
\end{table*}

\begin{table*}[]
\setcaptionmargin{0mm} \onelinecaptionstrue \captionstyle{normal}
\caption{Age in Gyr, $Y$, metallicity and abundances of chemical
elements determined in this study for Galactic globular clusters
as a result of modeling of their integrated-light spectra.
Superscript $^{\rm b}$ denotes the clusters belonging to the
bulge according to~Bica et al. (2016). We obtained two 
complete sets of parameters for NGC\,5986 and  NGC\,6441. For
NGC\,5946, NGC\,6235, NGC\,6522, NGC\,6569, the computation with two
isochrones yields similar results within the abundance measurement
errors (see Table~\ref{photMWGCs} and the text)}\label{tab_abund}
\medskip
\scriptsize
\begin{center}
\begin{tabular}{|l|c|c|c|r|r|r|r|r|r|r|r|}
\hline 
Parameter/   &  T$^{\rm our}_{\rm sp}$& Y$^{\rm our}_{\rm sp}$& [Fe/H] & [C/Fe]&[O/Fe]& [Na/Fe] &  [Mg/Fe]&[Ca/Fe]&[Ti/Fe]& [Cr/Fe]& [Mn/Fe]   \\
 Object     &   (Gyr)             &               &  (dex)   & (dex)   & (dex)  & (dex)     & (dex)     & (dex)  & (dex)  & (dex)   & (dex)     \\ 
\hline                                                                                                                                                                 
NGC\,1851  & 12.6           &   0.26        &  -1.60   &  0.00   & 0.40   & 0.35     & 0.51       & 0.35   & 0.45  &  0.10   &-0.10      \\ 
NGC\,2298  & 11.2           &   0.26        &  -2.20   &  0.15   & 0.50   & 0.40     & 0.40       & 0.45   & 0.20  &  0.30   & 0.20      \\ 
NGC\,2808  & 13.6           &   0.30        &  -1.30   & -0.35   & 0.10   & 0.45     & 0.20       & 0.15   & 0.05  &  0.00   & 0.00      \\ 
NGC\,3201  & 12.6           &   0.30        &  -1.50   & -0.10   & 0.30   & 0.20     & 0.20       & 0.10   & 0.10  &  0.00   & 0.00      \\ 
NGC\,5946  & 12.6,13.6      &   0.26        &  -1.95   &  0.15   & 0.55   & 0.45     & 0.45       & 0.35   & 0.35  &  0.15   &-0.05      \\ 
NGC\,5986  &      12.5      &   0.30        &  -1.75   & -0.10   & 0.30   & 0.00     & 0.25       & 0.15   & 0.10  &  0.20   & 0.20      \\
           & 11.2           &   0.30        &  -1.75   & -0.17   & 0.30   & 0.00     & 0.20       & 0.18   & 0.10  &  0.25   & 0.00      \\ 
NGC\,6171  & 10.0           &   0.26        &  -1.25   &  0.11   & 0.30   & 0.20     & 0.50       & 0.21   &-0.50  &  0.25   &-0.50      \\ 
NGC\,6218  & 12.6,13.6      &   0.26        &  -1.80   &  0.15   & 0.50   & -0.10     & 0.50       & 0.35   &-0.40  & -0.20   &-0.50      \\ 
NGC\,6235  &  12.6,13.6     &  0.26,0.30    &  -1.50   & -0.10   & 0.30   & 0.20     &-0.05       & 0.10   &-0.10  &  0.00   &-0.20      \\ 
 NGC\,6266 & 13.6           &   0.30        &  -1.50   & -0.10   & 0.30   & 0.20     & 0.35       & 0.25   &-0.30  &  0.30   &-0.40      \\ 
 NGC\,6284 & 12.6           &   0.30        &  -1.40   & -0.10   & 0.30   & 0.50     & 0.50       & 0.30   & 0.10  &  0.20   &-0.30      \\ 
 NGC\,6333 &  12.6          &   0.26        &  -2.30   & -0.07   & 0.30   & 0.20     & 0.30       & 0.45   & 0.30  &  0.20   & 0.00      \\ 
 NGC\,6342$^{\rm b}$ &  12.6      &   0.23        &  -0.90   &  0.10   & 0.30   & 0.20     & 0.50       & 0.00   &-0.50  & -0.10   &-0.30 \\  
 NGC\,6362$^{\rm b}$ &  13.6      &   0.30        &  -1.45   & -0.10   & 0.30   & 0.50     & 0.40       & 0.20   & 0.05  &  0.20   &-0.30 \\  
 NGC\,6441 &  10.0          &   0.26        &  -0.90    &  0.10   & 0.35   & 0.40      & 0.60        & 0.30    & 0.15  & 0.20     & -0.20      \\ %
           &  13.6          &  0.30         &  -0.95   &  0.00   & 0.30    & 0.20      & 0.40       & 0.30   & 0.10  & 0.00    & 0.00      \\ 
 NGC\,6522$^{\rm b}$ &  12.6      &  0.30,0.26    &  -1.45   & -0.10   & 0.30    & 0.20      & 0.60        & 0.20    & 0.30   & 0.10     & 0.2 \\   
 NGC\,6544 & 13.6           & 0.30          &  -1.80    & -0.20    & 0.30    & 0.20      & 0.20        & -0.10   & 0.10   & 0.00     & 0.00       \\ 
 NGC\,6569 & 13.6,11.2      & 0.23,0.26     &  -1.20    &  0.20    & 0.30    & 0.40      & 0.60        &  0.20   & 0.40   & 0.20     & 0.00       \\ 
 NGC\,6624$^{\rm b}$ & 13.6       & 0.23          &  -1.00    &  0.18   & 0.30    & 0.40      & 0.60        &  0.25  & 0.50   & 0.00     &-0.40  \\  
 NGC\,6626$^{\rm b}$ & 12.6       & 0.30          &  -1.50    & -0.30    & 0.30    & 0.45     & 0.30        &  0.20   & 0.35  & 0.10     &-0.50  \\  
 NGC\,6637$^{\rm b}$ & 12.6       &   0.23        &  -1.00   & -0.10   & 0.30   & 0.10     & 0.50       & 0.05   & 0.22  &  0.20   &-0.30 \\  
 NGC\,6638 & 13.6           & 0.30          &  -1.15   &  0.27   & 0.30    & 0.20      & 0.50        &  0.37  & 0.00   & 0.05    &-0.50       \\ 
 NGC\,6652 & 13.6           & 0.26          & -1.20     & 0.20     & 0.30    & 0.45     & 0.40        & 0.3    & -0.10  & -0.35   & -0.35     \\ 
 NGC\,6723$^{\rm b}$ & 12.6       & 0.26          & -1.40     & 0.05    & 0.30    & 0.50      & 0.40        & 0.35   & 0.15  & -0.20    & -0.30 \\  
 NGC\,6205 & 10.0           & 0.30          & -1.70     & -0.12   & 0.3    & 0.70      & 0.20        & 0.20    & 0.30   & -0.20    & 0.00       \\ 
 NGC\,7006 & 13.6           & 0.26          &  -1.8    & -0.07   & 0.45   & 0.40      & 0.27       & 0.20    & 0.05  & -0.25   & -0.05     \\
\hline                                    
\end{tabular}
\end{center}
\end{table*}

 \begin{figure}[]
 \setcaptionmargin{0mm} \onelinecaptionstrue \captionstyle{normal}
\includegraphics[scale=0.4,angle=-90]{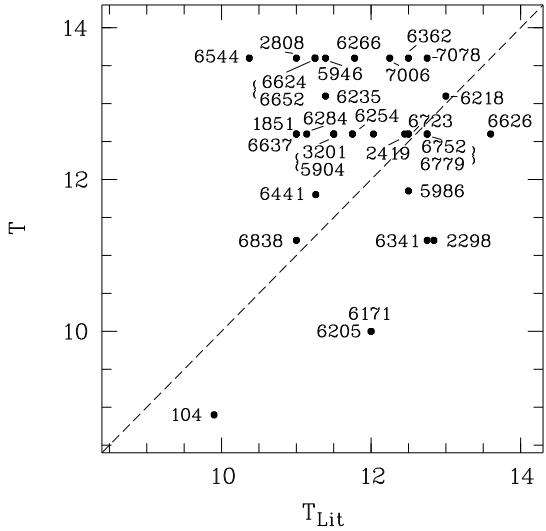} 
\caption{Comparison of the age estimates determined from the
analysis of their integrated-light spectra for Galactic globular
clusters with the values obtained in the literature using CMDs of
the clusters for all objects except NGC\,104 (see Sec.~\ref{comparlitAge}).}
 \label{fig_T}
\end{figure}

\begin{figure*}[]
 \setcaptionmargin{-5mm} \onelinecaptionstrue \captionstyle{normal}
\includegraphics[scale=0.5,angle=-90]{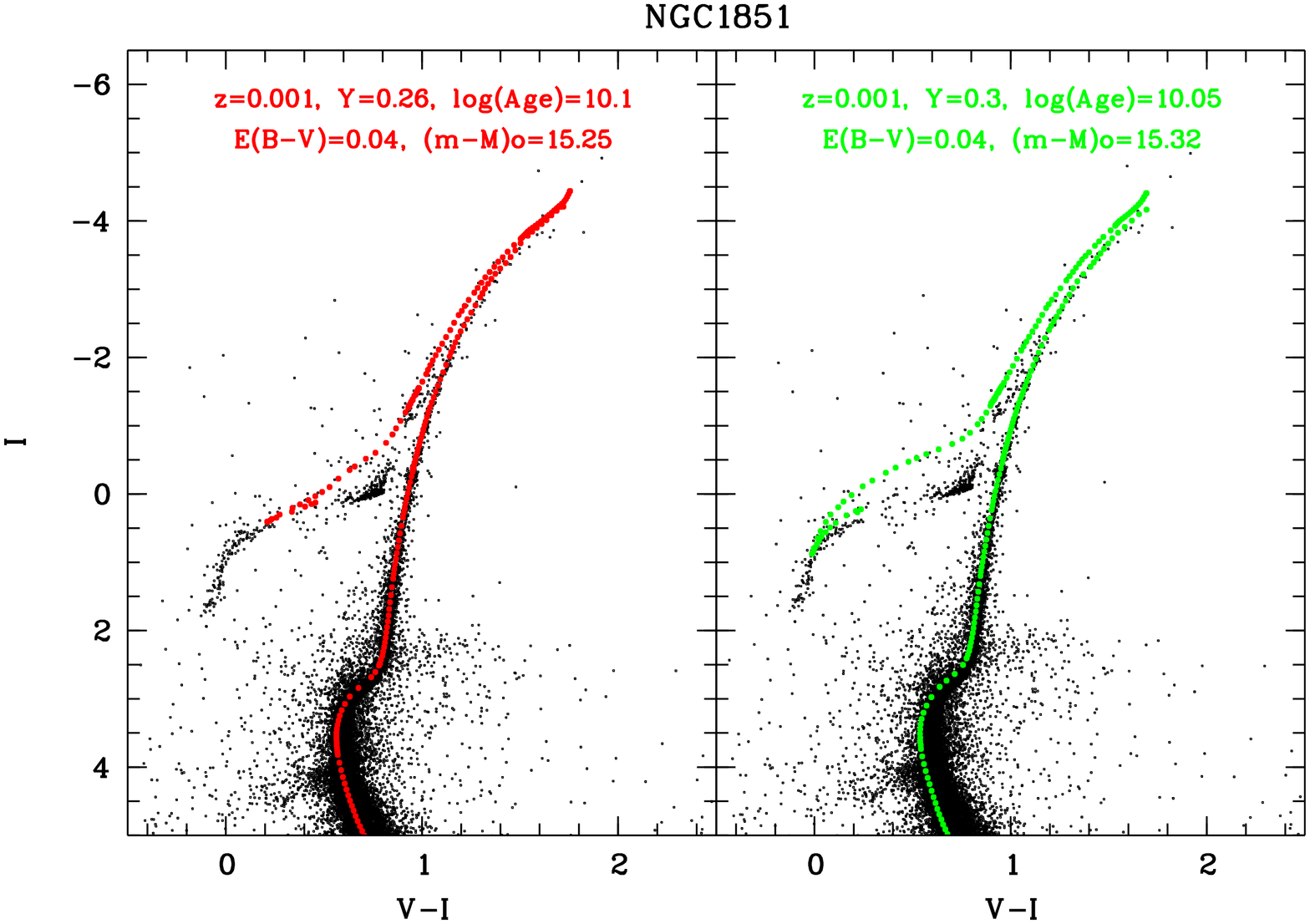} 
\includegraphics[scale=0.5,angle=-90]{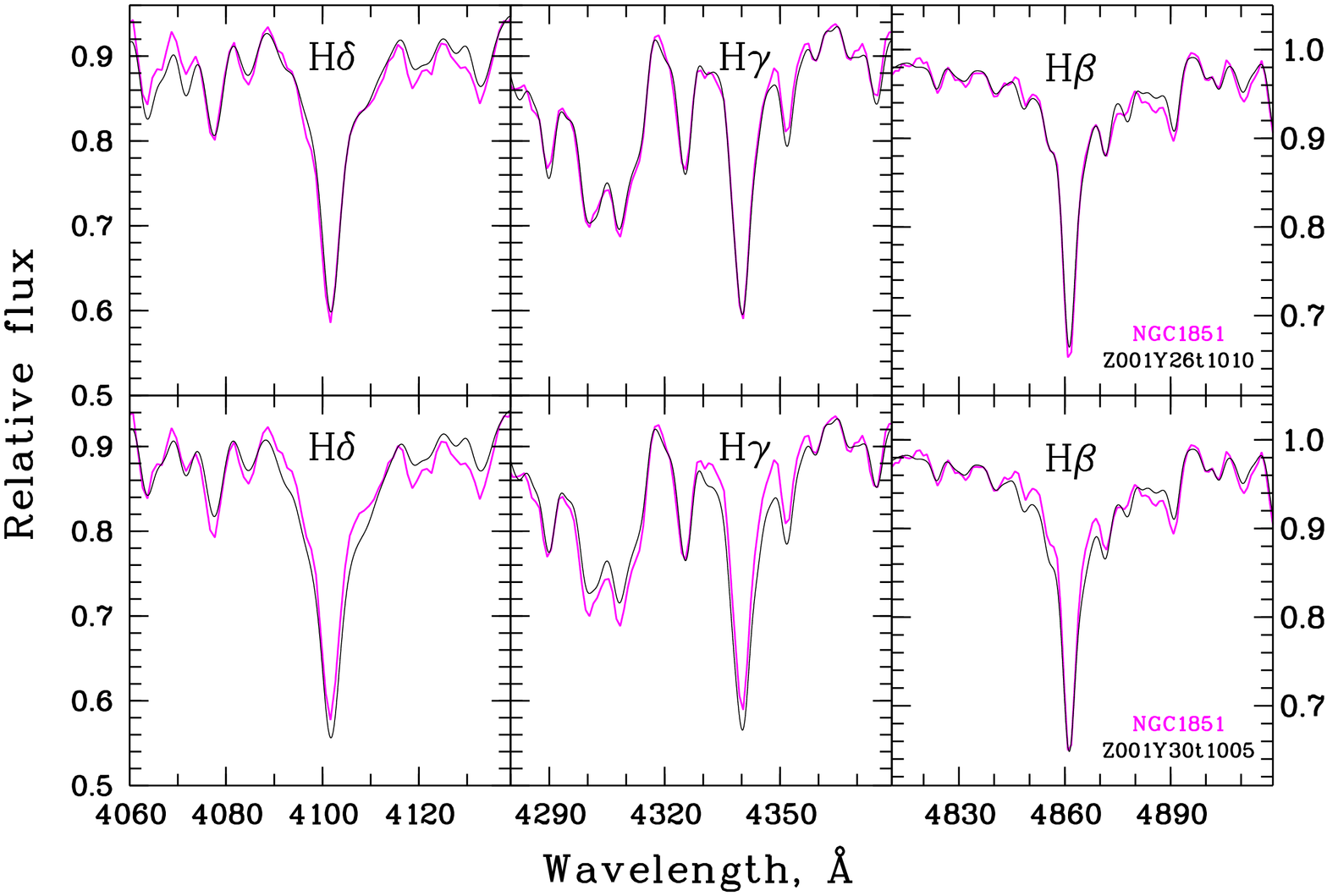} 
\caption{Comparison of the stellar photometry results for the
Galactic globular cluster NGC\,1851~(Sarajedini et al., 2007) with the
theoretical stellar evolutionary isochrones~(Bertelli et al., 2008) (top
panel). Comparison of the hydrogen lines in the observed spectrum
of NGC\,1851~(Schiavon et al., 2005) with the synthetic ones (black)
representing the integrated light of the cluster and computed
using the isochrones demonstrated in the top panel (bottom
panel).}
 \label{figN1851_1}
\end{figure*}

\begin{figure*}[]
 \setcaptionmargin{-5mm} \onelinecaptionstrue \captionstyle{normal}
\includegraphics[scale=0.5,angle=-90]{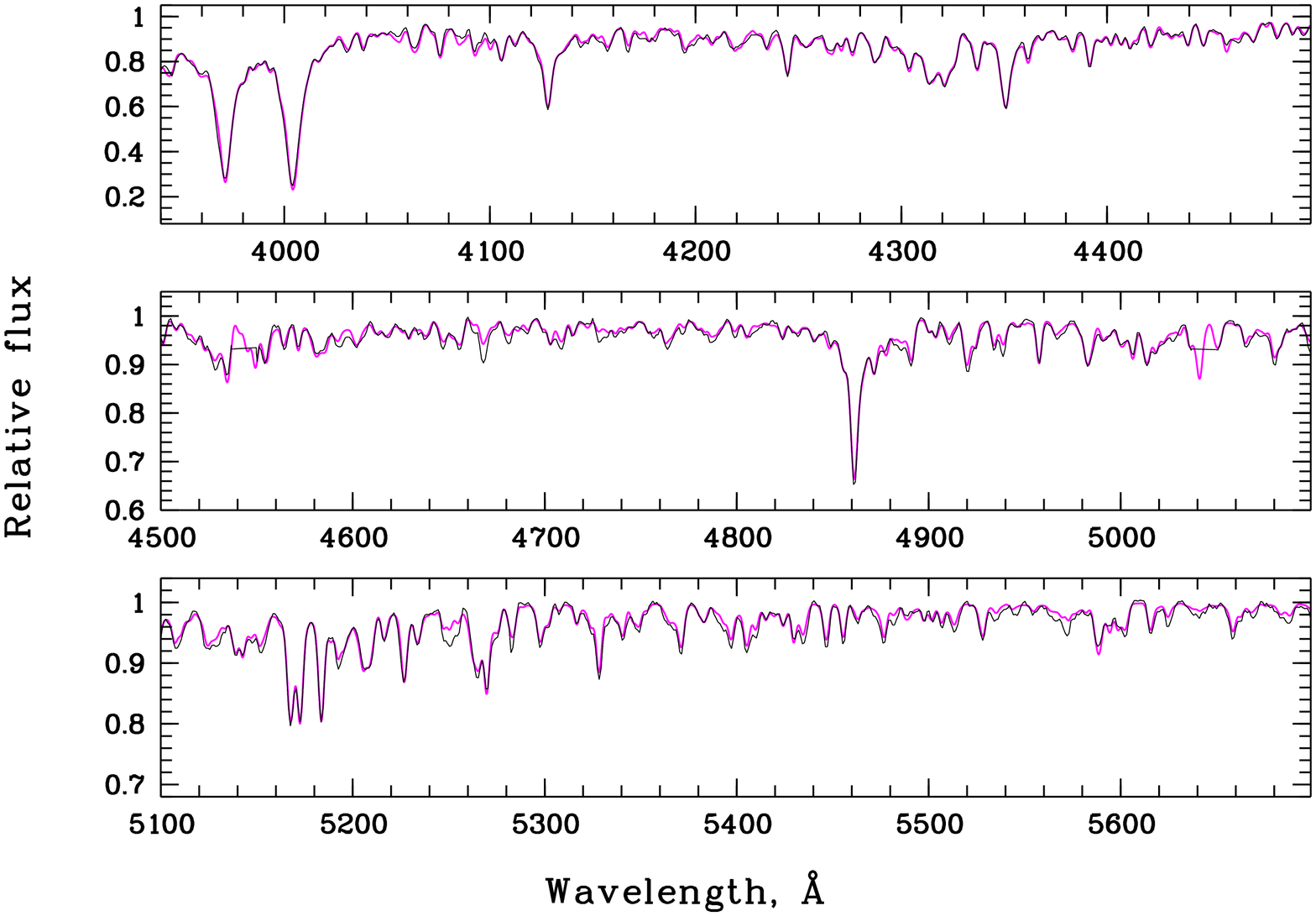} 
\caption{Comparison of the spectrum of
NGC\,1851 (black)~(Schiavon et al., 2005) with the synthetic spectrum of the
cluster (see text).}
 \label{figN1851_2}
\end{figure*}

\section{Results}
\label{results}
\subsection{Results of age, $Y$, $[Fe/H]$ and chemical abundances determination for Galactic globular clusters.}
Table \ref{photMWGCs} lists literature data for the objects of our
study obtained using their CMDs: metallicity, age, $Y$,
color-excess and apparent distance modulus, not corrected for the
Galactic extinction. The same table contains the determined in
this work \mbox{color-excess} values and the distance modules
corrected for the Galactic extinction. These data were obtained by
fitting the~Bertelli et al.~(2008) isochrones selected by studying the
integrated-light spectra of clusters and the distribution of stars on the
CMDs. The last column of the table shows the parameters of the
isochrones~(Bertelli et al.,2008) used in modeling the spectra of
clusters.

The ages,
helium mass fractions ($Y$), metallicities ([Fe/H]) and abundances
of chemical elements C, O, Na, Mg, Ca, Ti, Cr and Mn determined using 
the integrated-light spectra analysis for the test
sample of Galactic clusters are presented in Table~\ref{tab_abund}
and in Figs.~\ref{fig_Elem},~\ref{fig_abund} and~\ref{fig_T}.
Typical errors in the determination of the listed parameters using our
method and the integrated-light spectra with the signal-to-noise
ratio $S/N\sim100$ and the absence of the contribution from
Galactic field stars are the following: \mbox{$\sigma{\rm
[C/Fe]}\sim 0.2$}~dex, \mbox{$\sigma{\rm [O/Fe]}=0.3$}~dex, \mbox
{$\sigma{\rm [Na/Fe]}\sim 0.2$}~dex, $\sigma
\rm{[Mg/Fe]}\sim0.15$~dex, \linebreak \mbox{$\sigma{\rm
[Ca/Fe]}\sim 0.1$}~dex, $\sigma{\rm [Ti/Fe]}\sim 0.2$~dex,
\linebreak \mbox{$\sigma{\rm [Cr/Fe]}\sim 0.2$}~dex and
$\sigma{\rm [Mn/Fe]}\sim 0.2$~dex, \linebreak \mbox{$\sigma{\rm
[Fe/H]}\sim 0.1$}~dex. The accuracy of the age and $Y$
determination approximately corresponds to half the step of the
stellar evolution models used in these parameters
(see Khamidullina et al., 2014; Sharina et al. 2014, 2013, 2018, 2017). 
It should be noted that the oxygen lines are too weak
and not visible in the spectrum for the spectral range and
resolution that we use. However, the oxygen abundance affects the
formation of the molecular balance of CN and CH due to the
partial binding of carbon to the CO molecule. We usually set ${\rm
[O/Fe]}=0.3$ when started modeling the spectra. In some cases,
this content had to be slightly changed (Table~\ref{tab_abund})
so that the profiles of the molecular bands  CN and CH were
reproduced correctly. The intensities of these bands are
influenced by the lines of C, N, O and other elements.

In Figs.~\ref{fig_abund} and~\ref{fig_T}, we present a comparison
of the chemical abundances and ages determined for the clusters with
the corresponding literature values. We selected literature [Fe/H]
values mainly from the papers by~VandenBerg
et al.~(2013) and Harris~(1996). For NGC\,6624, NGC\,6341, NGC\,6637 and
NGC\,6652 [Fe/H], the values were taken from~Forbes and Bridges
(2010), and for
NGC\,6779,- from~Dotter et al. (2010). Literature abundances of chemical
elements were taken mainly from~Pritzl et al.~(2005) and Roediger
et al.~(2014). We found the missing data on the abundances
in the following papers:~Johnson et al.~(2017) for
NGC\,5986;~M{\'e}sz{\'a}ros et~al.(2015) for NGC\,6171 and
NGC\,6205;~Yong et al.~(2014) for NGC\,6266 and~Massari et al.~(2017) for
NGC\,6362.

\subsection{Comparison of the observed and synthetic\\ integrated-light spectra} \label{comparSec} 
The SAO ftp site\footnote{\url
{ftp://ftp.sao.ru/pub/sme/AnalILMWGCs/}} contains the figures of
the comparison of the observed  spectra of 26 Galactic globular
clusters from the paper by~Schiavon et al.~(2005) and the spectroscopic
OHP archive with the synthetic spectra calculated using models of
stellar atmospheres. Also, it includes a comparison of literature stellar
photometry results in these globular
clusters~(Sarajedini et al., 2007; Piotto et al., 2002) with the stellar
evolutionary isochrones selected based on the best 
description of the hydrogen lines in the observed spectra. 
Here, we present the aforementioned comparison for only one object.
Figure~\ref{figN1851_1} (top panel) shows the comparison of the
stellar photometry results~(Sarajedini et al., 2007) with two
isochrones~(Bertelli et al., 2008) selected based on the best fitting of
the hydrogen lines by the model ones. Both isochrones describe the
distribution of stars on the CMD correctly, except for the HB stage.
The horizontal branch is not described precisely in either case.
The comparison of the hydrogen line profiles in
the observed and synthetic spectra calculated using two isochrones
shows that an isochrone with the absence of the extremely blue HB stars is
more suitable: \mbox{$Z = 0.001$}, \mbox{$Y = 0.26$}, \mbox{$\log
T = 10.10$}. It can be assumed that the integrated spectrum of the
cluster contains a relatively small number of such objects.
Figure~\ref{figN1851_2} shows a complete comparison of the
observed and synthetic spectra. The model spectrum was computed
using the isochrone \mbox{$Z = 0.001$}, \mbox{$Y = 0.26$},
\mbox{$\log T = 10.10$} and abundances from Table~\ref{tab_abund}.
Note, that the spectra of globular clusters NGC\,5946, NGC\,6235,
NGC\,6333, NGC\,6342, NGC\,6544, NGC\,6569 from~Schiavon et al. (2005)
have rather low $S/N$. This can be seen in the figures of their
comparison with the model spectra presented on the SAO ftp site.

Figure~\ref{fig_Elem} shows the estimates of the abundances
derived by us together with the data from our previous works for
the Galactic (NCG\,104, NGC\,1904, NGC\,2419, NGC\,2808,
NGC\,3201, NGC\,5286, \linebreak NGC\,5904, NGC\,6121, NGC\,6205,
NGC\,6254, \linebreak NGC\,6341, NGC\,6752, NGC\,6779, NGC\,6838,
\linebreak NGC\,7089) and extragalactic (Mayall\,II, SD9-GC7,
MGC\,1, Bol\,298, GC-KKs3, GC-E269-66, CBF\,28, CBF\,98)
clusters: Khamidullina et al. (2014); Sharina
et al. (2014); Sharina and Shimansky (2020); Sharina
et al. (2013, 2018, 2017). The objects are grouped according to their
[Fe/H] values. Typical errors of the abundances are shown in the
top left panel. This panel summarizes the data for the objects with
\mbox{${\rm [Fe/H]} \sim -1$}~dex. The middle top panel of
Fig.~\ref{fig_Elem} shows abundances of chemical elements for the
clusters with ${\rm [Fe/H]} \sim -2$~dex. Other panels of the
figure demonstrate the objects with \mbox{${\rm [Fe/H]} = -1.5
\pm 0.25$}~dex.

\subsection{Comparison of our and literature age estimates}
\label{comparlitAge} Comparison of our and literature data was
carried out for the full sample of 40 clusters, which is described
in Section~2. We used the available literature estimates by VandenBerg
et al. (2013) and the absent values were taken from the
papers by Kruijssen et al. (2019) and Testa et al. (2001). We used the age 
$9.9\pm0.7$~Gyr for NGC\,104 determined
by Hansen et al. (2013) during the investigation of its white dwarf sequence.
It should be noted that literature data about the absolute age of
NGC\,104 are quite different (Brogaard et al., 2017). When comparing
our and literature estimates of age, we considered all literature
age values greater than the age of the Universe to be equal to
13.6~billion years. We did not manage to find in the literature  
absolute age values for five sample clusters: NGC\,6229,
NGC\,6333, NGC\,6522, NGC\,6569, NGC\,6638. A comparison of our
absolute age estimates with the corresponding literature values
(Fig.~\ref{fig_T}) shows that the moduli of the differences
between them are more than 1.4~billion years for 14 objects, that
is, for 40\% of the sample. 
Let us briefly consider the reasons of
these deviations and, in general, the reasons behind the significant
differences in the estimates of the absolute ages of Galactic globular
clusters by different authors.

More than half of the 14 objects are located close to the plane of
the Galaxy and have a significant color excess \mbox{$E(B-V)\ge
0.3$} (Harris, 1996; Schlegel et al., 1998). In such cases, the quality of the
spectrum is determined not only by the brightness of the object
itself but also by the influence of a large number of background
stars. They contribute both to the total spectrum of the cluster
and to the total spectrum of background stars, which is subtracted
from the spectrum of the cluster. Thus, due to significant
inhomogeneities in the field, the influence on the spectrum
of the field stars can be overestimated or underestimated.
The intensities and shapes of the lines of hydrogen and of various
chemical elements in the integrated-light spectra may be randomly
distorted. Therefore, we meet inevitable difficulties in the
analysis of the spectra. Moreover, Schiavon et al. (2005) noted that
the spectra of clusters at low galactic latitudes can be distorted
by weak emission lines of the galactic background. Note that the
described difficulties in obtaining and analyzing
integrated-light spectra are significant for the objects of our
Galaxy, which are large in projection onto the celestial sphere,
that is, the probability of the field pollution of their
spectra is high.

Furthermore, for nine of the aforementioned 14 objects,
the metallicity is \mbox{${\rm [Fe/H]} \le -1.4$~dex}. It has long
been established that the \mbox{well-known}
\mbox{age-metallicity} degeneracy problem is more difficult to
solve for low metallicities (see e.g.~VandenBerg et al. (2013),
Pietrinferni et al. (2013)). This is mainly due to the negligible
changes in the observed parameters that occur in this case with
varying the metallicities. In addition, let us note that the
results of age and metallicity determination using
\mbox{color--magnitude} diagrams depend on the adopted oxygen
abundance in stellar population
models~(Denissenkov et al., 2017; VandenBerg et al., 2013). The situation is
complicated by Na--O anticorrelations in globular clusters.

For seven objects in the sample, the absolute values of the
differences between the literature and our age estimates exceed 2~Gyr:
$${\rm NGC}\,2808\!:~|\Delta(T)|\!=\!2.6~{\rm Gyr}, E(B\!-\!V)\!=\!0.23~{\rm mag},$$\\[-60pt]
$${\rm NGC}\,5946\!:~|\Delta(T)|\!=\!2.2~{\rm Gyr}, E(B\!-\!V)\!=\!0.54~{\rm mag},$$\\[-60pt]
$${\rm NGC}\,6171\!:~|\Delta(T)|\!=\!2.0~{\rm Gyr}, E(B\!-\!V)\!=\!0.33~{\rm mag},$$\\[-60pt]
$${\rm NGC}\,6205\!:~|\Delta(T)|\!=\!2.0~{\rm Gyr}, E(B\!-\!V)\!=\!0.02~{\rm mag},$$\\[-60pt]
$${\rm NGC}\,6544\!:~|\Delta(T)|\!=\!3.23~{\rm Gyr}, E(B\!-\!V)\!=\!0.73~{\rm mag},$$\\[-60pt]
$${\rm NGC}\,6624\!:~|\Delta(T)|\!=\!2.35~{\rm Gyr}, E(B\!-\!V)\!=\!0.28~{\rm mag},$$\\[-60pt]
$${\rm NGC}\,6652\!:~|\Delta(T)|\!=\!2.35~{\rm Gyr}, E(B\!-\!V)\!=\!0.09~{\rm mag}.$$\\[-45pt]

As it was mentioned in Section~\ref{comparSec}, NGC\,5946 and
NGC\,6544 have low $S/N$ ratio in their spectra. Color excess
values $E(B-V)$ for all objects, except NGC\,2808, NGC\,6205 and NGC\,6652,
are are greater or equal to $0.3$~(Harris 1996; Schlegel et al., 1998).
Probably, the spectra of these objects were contributed by
background stars.

Let us analyze the reasons for the discrepancy between our and
literature age data for three clusters with low \mbox{$ E (B-V) =
0.09 $}. NGC\,2808 is an unusual cluster, whose stars show
significant variations in the helium content at a constant
nitrogen abundance, similar to that found in field stars
(Cabrera-Ziri et al.~(2019) and references therein). Significant scatter
of colors and luminosities of stars relative to the inscribed
evolutionary sequences can be seen on the CMD of the
cluster\footnote{\url
{ftp://ftp.sao.ru/pub/sme/AnalILMWGCs/NGC1851_2298_2808_3201.pdf}}
plotted according to the stellar photometry results
of Sarajedini et al. (2007). Our analysis of the spectrum of the cluster
reveals the metallicity lower than it follows from the analysis of
the object's CMD. We did not manage to describe the horizontal
branch of NGC\,2808  by isochrones (Bertelli et al., 2008). It is wide,
extremely extended to the blue side with a large number of faint
stars on the blue end and bright stars on the red end. More
detailed spectroscopic studies, determination of the helium content of
individual stars, and more sophisticated stellar evolution models
are needed to better understand the properties of the integrated
light of NGC\,2808.

NGC\,6205 (M\,13) is a nearby, bright and well-studied object.
Narrow evolutionary sequences and extremely blue horizontal branch
can be seen at the cluster CMD\footnote{\url
{ftp://ftp.sao.ru/pub/sme/AnalILMWGCs/NGC6638_6652_6723_6205_7006.pdf}}
according to the stellar photometry results (Sarajedini et al., 2007).
The specific helium content varies significantly among cluster
stars~(Denissenkov et al., 2017). These authors observed unusually high
periods of the RR\,Lyr stars in the object. We did not manage to
describe the integrated-light spectrum of NGC\,6205 with
the~Bertelli et al.~(2008) models. The spectrum of NGC\,6205 we used
 was obtained from observations in OHP in one spectroscopic slit
position during 300~sec. and is not representative of the total
stellar population of the cluster. The younger age derived by us
likely indicates that quite a few blue hot horizontal branch fell
into the slit during observations.

In the article by Sharina et al. (2020), we analyzed
the spectrum of NGC\,6652 from~Schiavon et al.~(2005). In that study,
the observed spectrum was compared with synthetic spectra
calculated using not only the~Bertelli et al.~(2008) isochrones, but
also the Teramo group isochrones (Pietrinferni et al., 2013) as well
as using different stellar mass functions. There is still no
consensus in the literature about the age of NGC\,6652 and its
association with the Galactic substructures~(e.g. Sharina et al., 2020)
and references therein). Our analysis of the cluster spectrum
from~Schiavon et al.~(2005) argues in favor of the presence of blue
horizontal branch stars in NGC\,6652. Our results indicate that
the stellar population of the object can be characterized by the
chemical composition and age typical of the Galactic bulge
globular clusters. The study of the shapes and depths of the
hydrogen lines in combination with the analysis of the
distribution of stars on the cluster CMD has led to the conclusion
that NGC\,6652 is older than it was accepted in the literature:
13.6~Gyr instead of 11.7~Gyr~(Chaboyer et al., 2000), or
11.25~Gyr~(Van den Berg et al., 2013).

\subsection{Comparison of the abundances with literature data}
\label{comparlit}

Figure~\ref{fig_abund} (left panels) represents
the differences between the elemental abundances determined in
this study and the corresponding literature values from high-resolution
spectroscopic studies of the brightest cluster stars.
The mean difference and the standard error of the mean are given
in each panel legend. There are no significant systematic
differences, except for the cases of [Fe/H] and [C/Fe]. It can be
seen in Fig.~\ref{fig_abund} that [Fe/H] values determined by us
are on average 0.2~dex less than the literature ones. This is a
distinctive feature of our method in it's current
state. A possible reason may be an overestimation of the $\xi_{\rm
turb}$ values used by us, which affect the line intensities in the
synthetic spectra and the abundances of chemical elements.
The overestimation of $\xi_{\rm turb}$ lowers the abundances of
chemical elements, primarily iron, whose lines prevail in the
spectra. The question needs further investigation. The reasons
for the systematic differences of [C/Fe] will be discussed later
in this section.

The deviations of [Fe/H] values from literature values are larger than for other sample
clusters for several objects with \linebreak \mbox{${\rm [Fe/H]}$} in the range
\mbox{$[-1.8; -1]$}~dex. These are NGC\,1851, NGC\,3201, NGC\,6218,
NGC\,6333, NGC\,6362 and NGC\,5946. As it was mentioned earlier in
this paper, NGC\,6333 and NGC\,5946 have low $S/N$ in their
spectra. Let us consider the cases of NGC\,1851, NGC\,6218,
NGC\,6362 and NGC\,3201 in more detail.

The deviation of the [Fe/H] value for NGC\,1851, and
probably for many other clusters in the sample, is likely due to
the fact that we use scaled solar isochrones of stellar evolution
and models of stellar atmospheres. Elemental abundance ratios
differ significantly from those of the Sun in NGC\,1851, the
globular cluster Omega\,Centauri and several other galactic clusters.
NGC\,1851 contains so called ``anomalous'' second generation stars
(see e.g.~Simpson et al., 2017) and references therein). Variations of
C, N, O elemental abundances in them correlate with the abundances of
$s$-process elements and with the metallicity [Fe/H]. Typically,
two stellar populations exist in such ``anomalous'' clusters:
\mbox{$s$-element/Fe-rich} and \mbox{$s$-element/Fe-poor}. Four
stellar populations were found in NGC\,1851~(Simpson et al. 2017 and
references therein). A more detailed investigation of the
\mbox{integrated-light} spectrum of this cluster would be desirable.

The large deviation of [Fe/H] from the literature value for
NGC\,6218 is apparently due to two main reasons. First, the Poisson
noise in the spectrum is quite large\footnote{\url
{ftp://ftp.sao.ru/pub/sme/AnalILMWGCs/NGC5946_5986_6171_6218.pdf}},
especially in its red part, the lines in which are mainly used for
metallicity determination. The cluster is located at a low
Galactic latitude. Therefore, the background subtraction result
may be imperfect. Second, the cluster is unusual, with an extremely
blue horizontal branch. It can be seen from the figures presented
on the \mbox{ftp-site} that the isochrones by Bertelli et al.
(2008) selected for modelling
the spectrum do not describe perfectly either the spectrum or the
CMD of the cluster. Probably, the reason is in the significantly
non-solar elemental abundance ratios for NGC\,6218 (see
e.g.~Mishenina et al., 2003).

During the observations of NGC\,6362, Schiavon et al.~(2005) encountered problems
with the line widths in the calibration spectrum of the arc lamp
(Section~3.5.2 in~Schiavon et al., 2005). Therefore, the authors
advised to use this spectrum only for low-resolution analysis.

NGC\,3201 has a loose structure and is located at a low Galactic
latitude. Schiavon et al.~(2005) discovered the emission line
[O\,II]~$\lambda 3727$~\AA\ in its integrated-light spectrum and
suggested that the Balmer series lines can also be distorted by
weak H\,II-emission of the galactic background. We have managed to
fit the hydrogen lines in the spectrum of NGC\,3201 by the model
ones. However, the $S/N$ ratio in the spectrum is quite
low\footnote{\url
{ftp://ftp.sao.ru/pub/sme/AnalILMWGCs/NGC1851_2298_2808_3201.pdf}},
and this fact also influenced the fitting results.

In the following, we will continue the comparison of our and
literature results. Figure~\ref{fig_abund} (right panels)
demonstrates that the abundances of elements determined by us are
consistent within the errors with the literature values obtained by
studying the integrated spectra of the clusters. For some objects,
there are systematic differences between our and literature
data. For example, the estimates by~Colucci et al.~(2017) of
[Mg/Fe] abundances in four objects common with our sample are
lower on average by 0.2~dex than our estimates. The
estimates by Larsen et al.~(2017) of [Mg/Fe] abundances in NGC\,7078,
NGC\,6254, and NGC\,6752 are somewhat higher than our estimates.

The reasons for the differences may lie not only in the
dissimilarities between the applied methods, but also in the
contribution of various stars into the integrated spectra analyzed
by the authors. The spectra were obtained using different telescopes and
spectrographs. It is possible that background stars fall into the
integrated spectra. Conroy et al.~(2018) uses the spectra 
from Schiavon et al.~(2005), that is, the same data that
we analyzed. The [Mg/Fe] estimates made by Conroy et al.~(2018) are on
average consistent with our results. The variance of the
differences is 0.16~dex and corresponds to the
average abundance estimation error. 
Let us recall that some spectra
of objects from~Schiavon et al.~(2005) were obtained with a relatively
low signal-to-noise ratio. A number of clusters are close to the Galactic
plane. Therefore, despite of a careful
background subtraction, their spectra can
be significantly changed by background stars.

Note the difference between the [C/Fe] values obtained using the
integrated spectra of clusters and the spectra of their brightest
stars (Fig.~\ref{fig_abund}, two bottom panels). While the average
difference between our estimates and the~Conroy et al.~(2018) data is
$-0.066$~dex, the difference between our  [C/Fe] for the clusters
and the mean [C/Fe]  of the brightest stars in the clusters is on
average  $+0.38$~dex. We interpret the latter as the effect of a
change in the chemical composition of the atmospheres of stars
in the course of their evolution (Kraft,~1994).

\section{Elemental abundances in globular clusters depending on metallicity and comparison with chemical evolution models}
\begin{figure*}[]
 \setcaptionmargin{5mm} \onelinecaptionstrue \captionstyle{normal}
\begin{tabular}{p{0.5\textwidth}p{0.5\textwidth}}
 \includegraphics[scale=0.27,angle=-90]{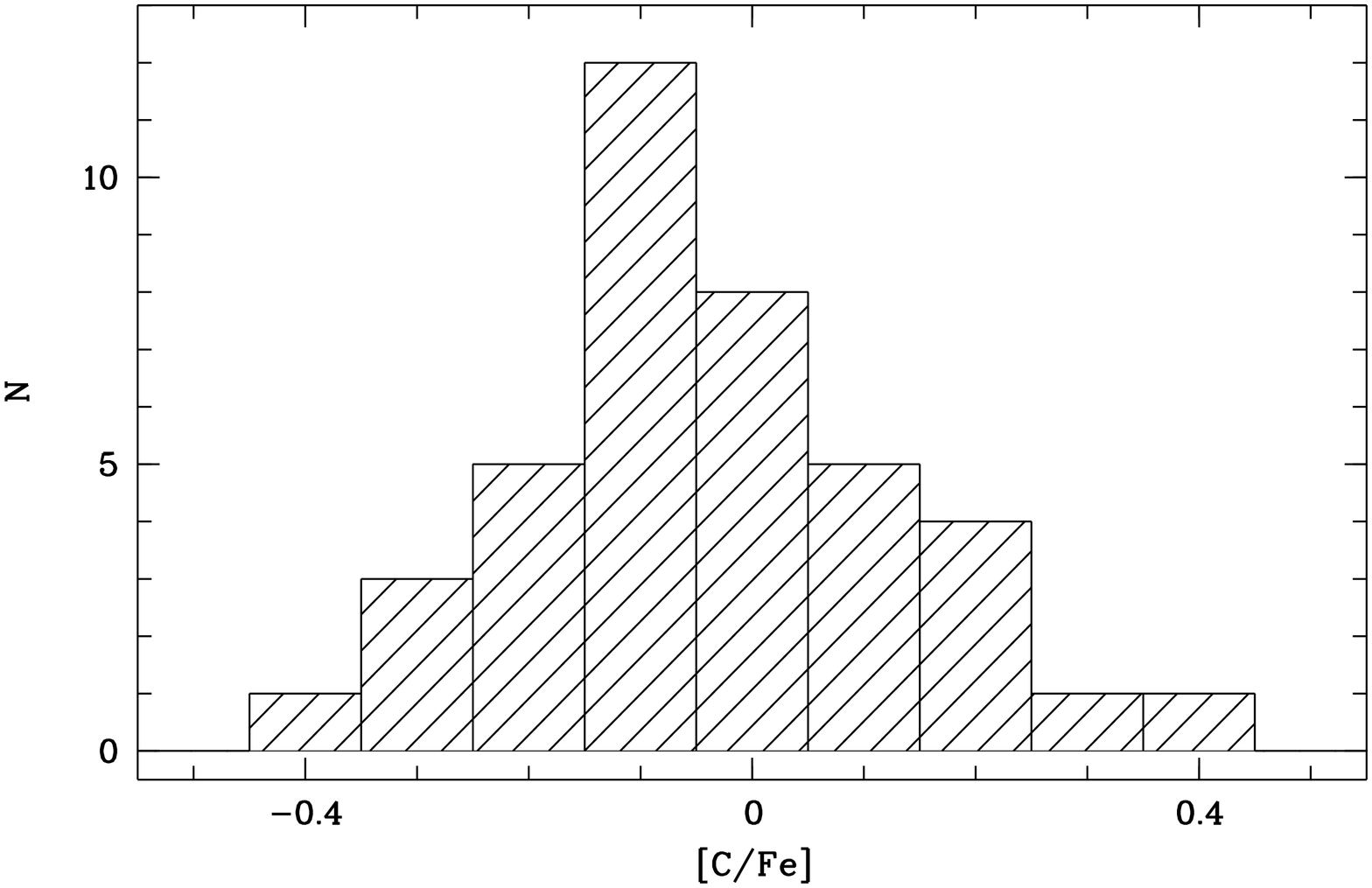} &
\includegraphics[scale=0.27,angle=-90]{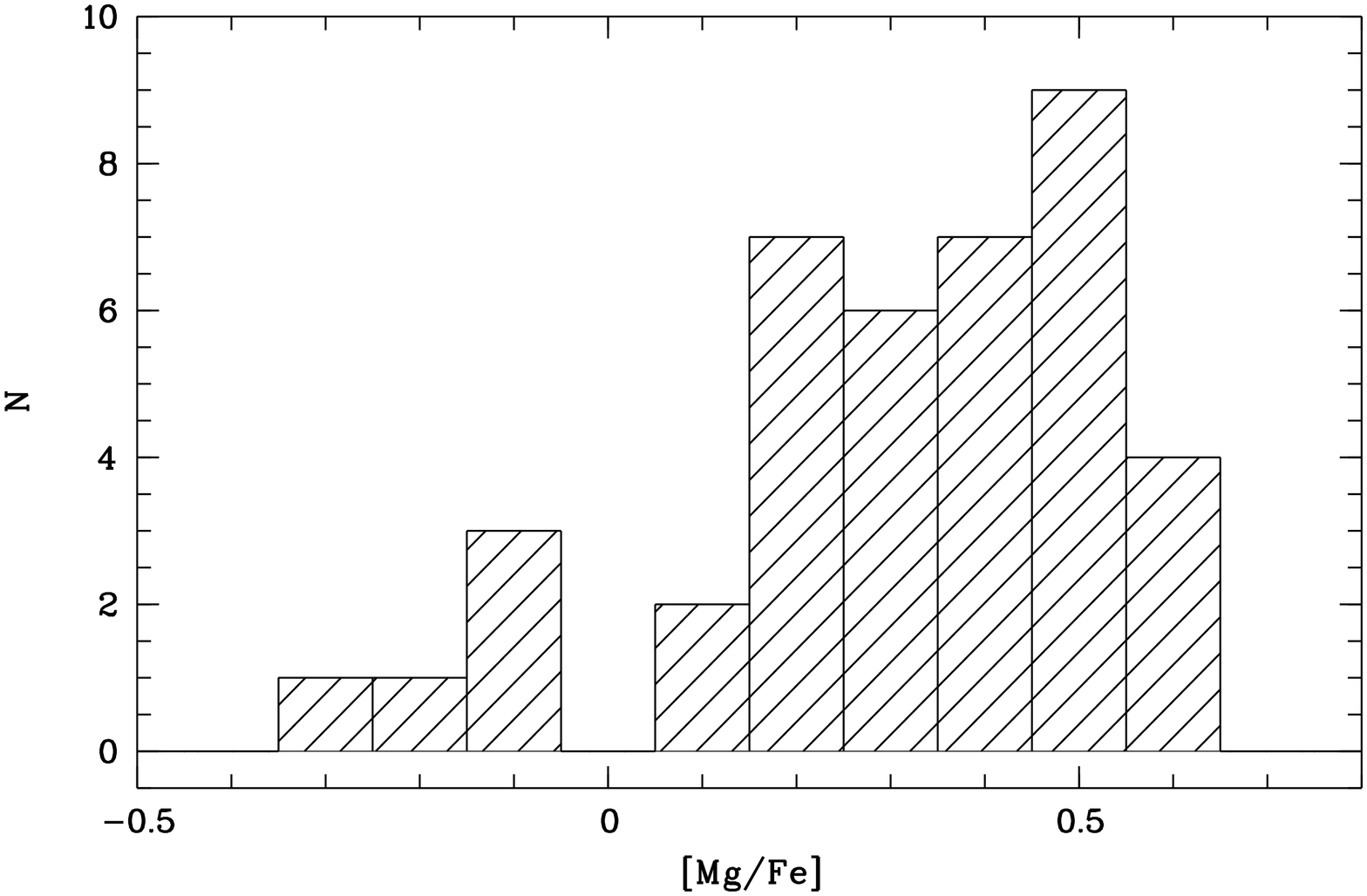} \\   %
\includegraphics[scale=0.27,angle=-90]{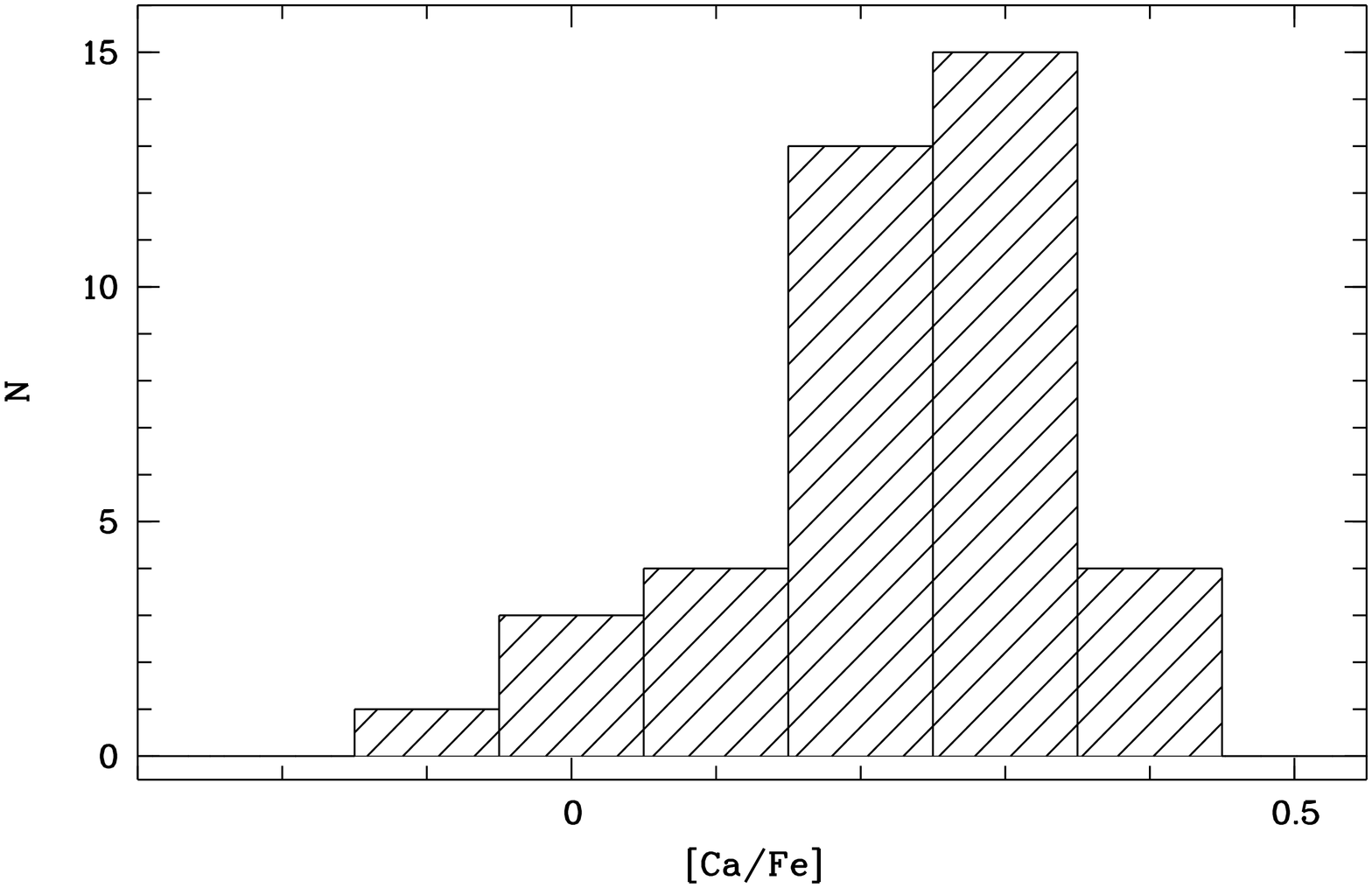} &
\includegraphics[scale=0.27,angle=-90]{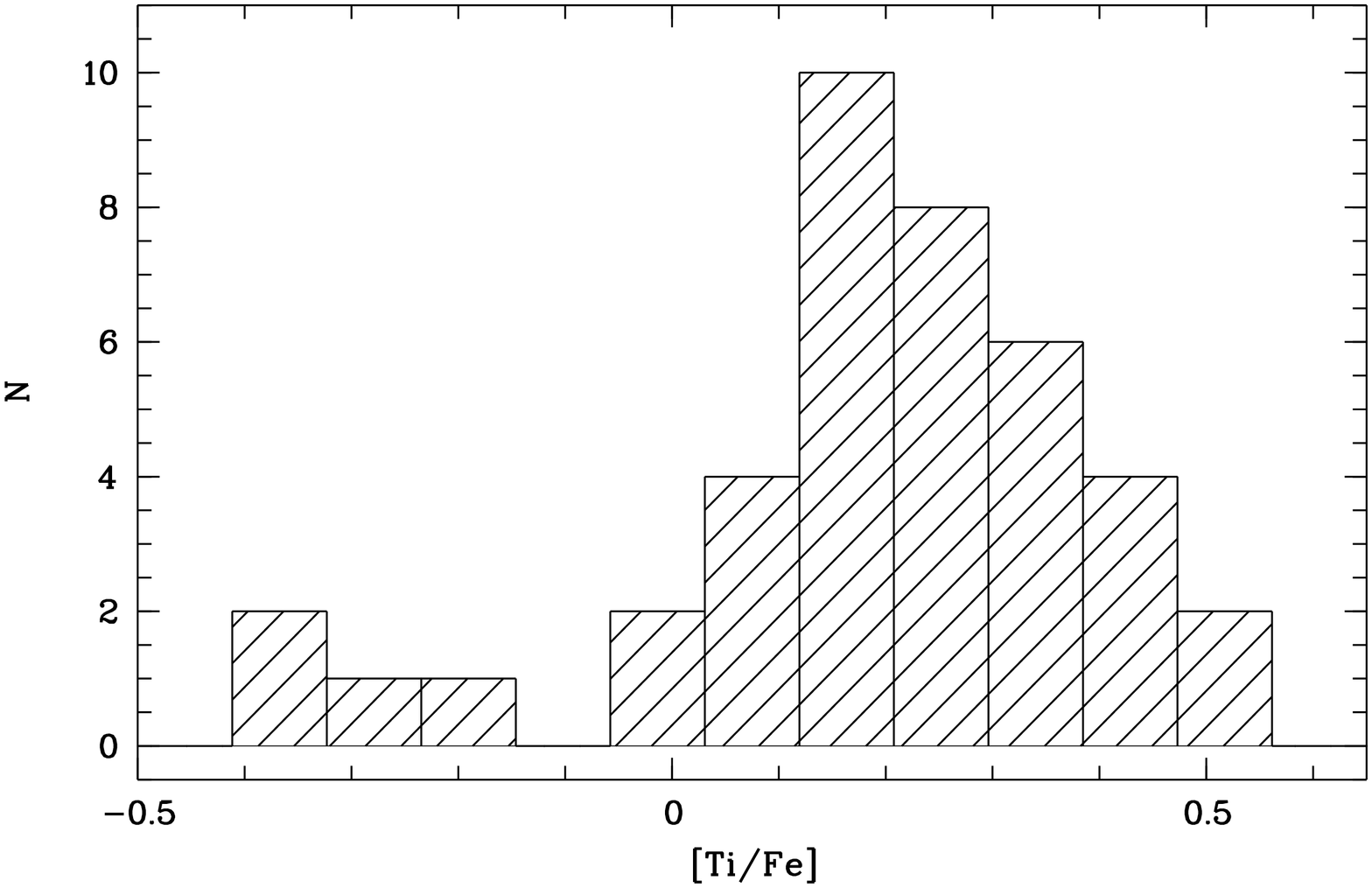} \\
 \includegraphics[scale=0.27,angle=-90]{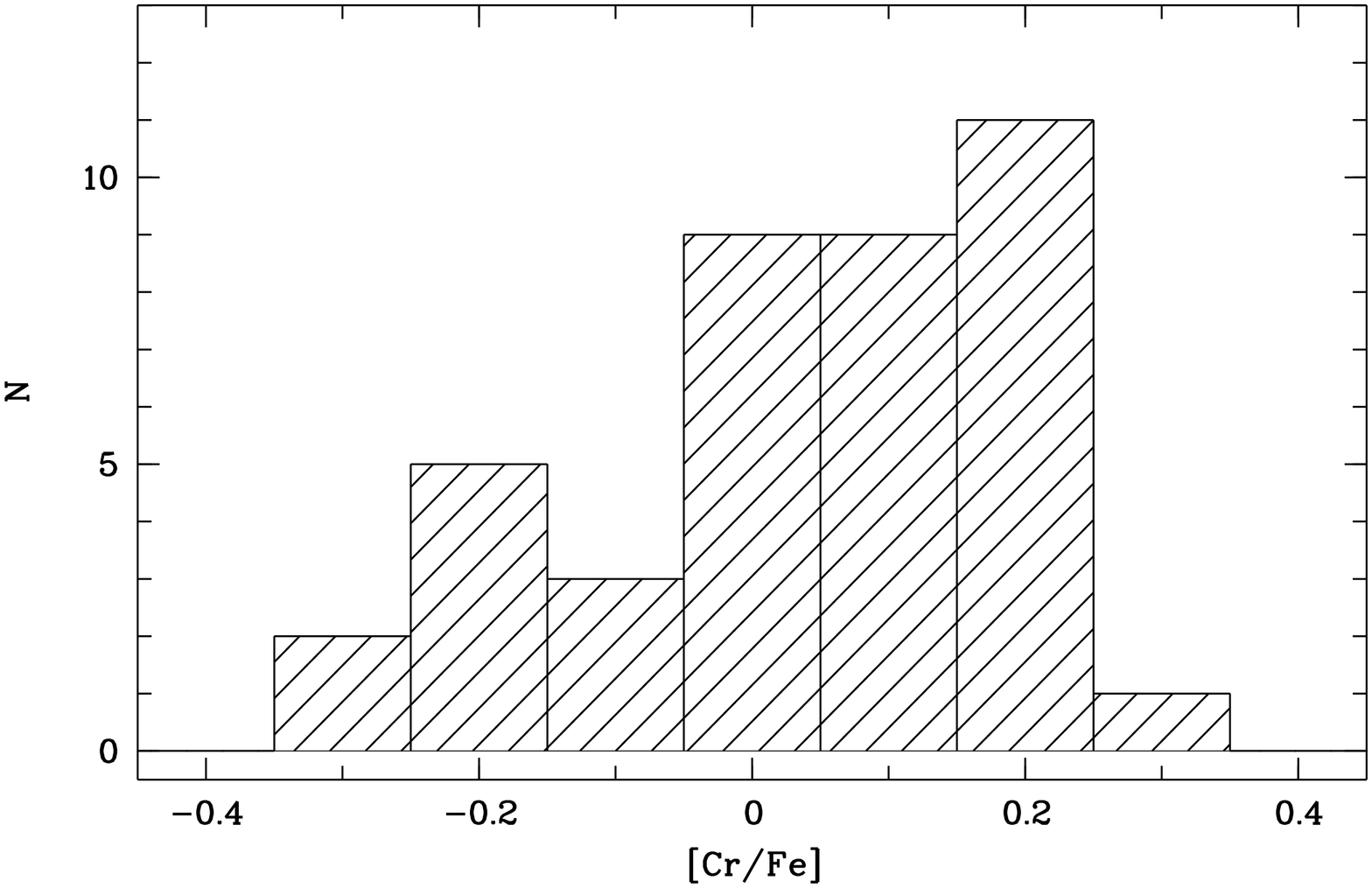} &
 \includegraphics[scale=0.27,angle=-90]{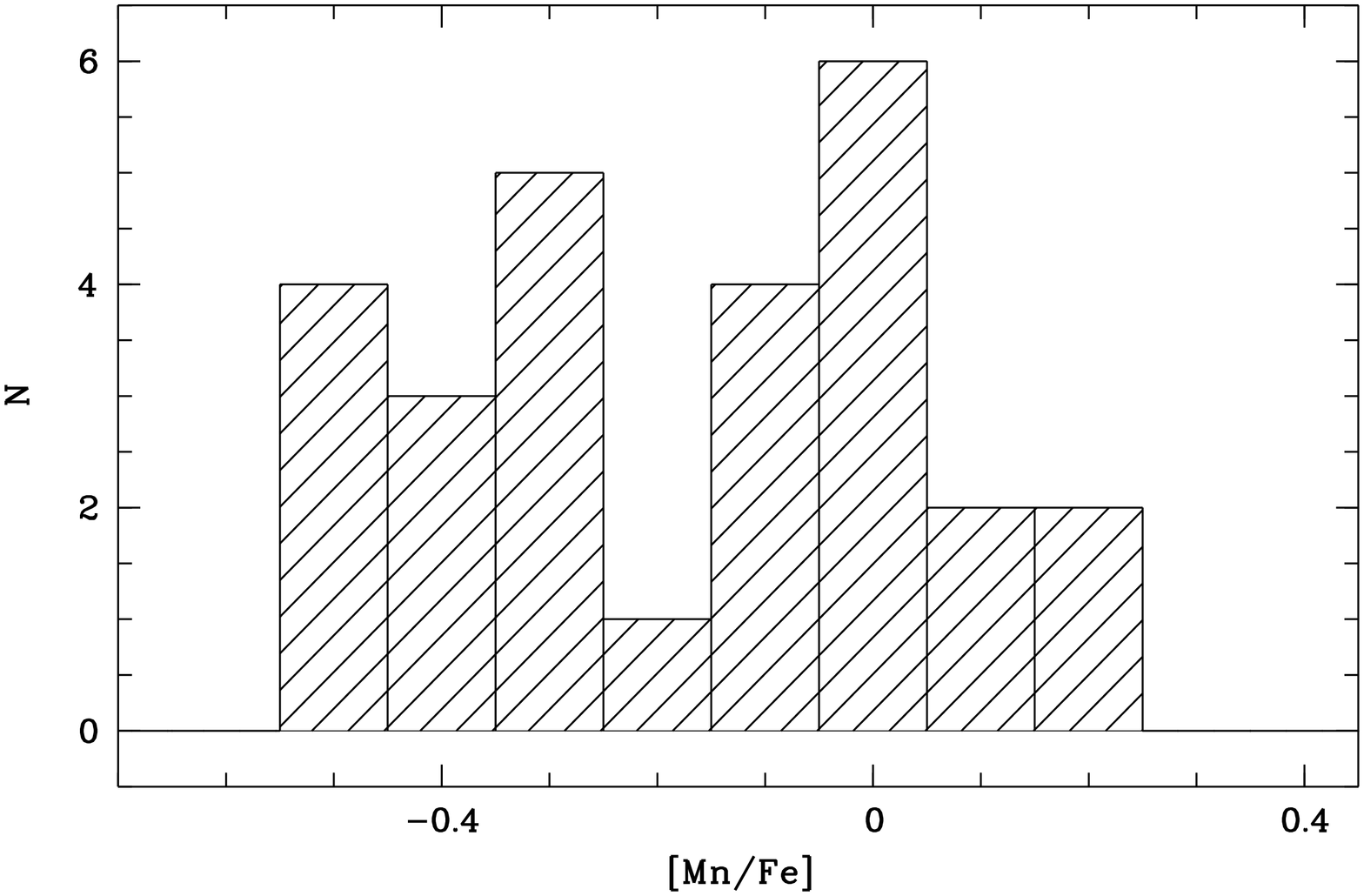} \\
 \end{tabular}
 \caption{Distributions of the determined abundances of chemical elements for 40 Galactic globular clusters (see text).}
 \label{fig_Hist}
\end{figure*}

The distribution of the abundances of chemical elements for 40
Galactic globular clusters are shown in Fig.~\ref{fig_Hist}. The
mean abundance values and the standard deviations for 40 objects
are the following: ${\rm [C\!/Fe]}\!=\!-0.025\!\pm\!0.17\,{\rm dex},$ \\ [-30pt] 
$${\rm [Mg\!/Fe]}\!=\!0.32\!\pm\!0.22\,{\rm dex}, {\rm [Ca/Fe]}=0.22\!\pm\!0.12\,{\rm dex}, $$\\[-60pt] 
$${\rm [Ti\!/Fe]}=0.14\!\pm\!0.24\,{\rm dex}, {\rm [Cr\!/Fe]}\!=\!0.05\!\pm\!0.16\,{\rm dex}, $$\\[-60pt] 
$${\rm [Mn\!/Fe]}\!=\!-0.17\!\pm\!0.22\,{\rm dex},{\rm [Na\!/Fe]}\!=\!0.3\pm0.2{\rm dex}.$$ 
It can be seen in~Fig.\ref{fig_Elem}
that, in general, in three groups of clusters with different
metallicities, the distributions of the abundances look similar.

The majority of the sample objects have \linebreak \mbox
{${\rm[Fe/H]}\le-1.0$}~dex and apparently belong to the Galactic halo 
divided by the internal and the external components
according to~\citet{Carretta}. The problems of separation of
clusters into subsystems are discussed e.g. by~Marsakov et al. (2019).
A part of our sample clusters belongs to the Galactic
bulge~(Bica et al.,~2016): NGC\,6342 (\mbox{${\rm [Fe/H]}_{\rm
our}=-0.9$}, ${\rm [Mg/Fe]}_{\rm our}=0.5$), NGC\,6522 ($-1.45$,
$0.6$), NGC\,6624 ($-1.0$, $0.6$), NGC\,6626 ($-1.5$, $0.3$),
NGC\,6637 ($-1.0$, $0.5$), NGC\,6723 ($-1.4$, $0.4$). Some of our
sample objects, apparently, fell into the bulge region from the
outer parts of the Galaxy~(Bica et al.,~2016): NGC\,6333 (\!\mbox
{${\rm [Fe\!/\!H]}_{\rm our}\!\!=\!-2.3$}, ${\rm [Mg/Fe]}_{\rm
our}=0.3$), NGC\,6171 ($-1.25$, $0.5$), \linebreak NGC\,6235
(\!$-1.5$, $-0.05$\!), NGC\,6441 (\!$-0.9$, $[0.4; 0.6]$\!),
NGC\,6544 ($-1.8$, $0.2$), NGC\,6569 ($-1.2$, $0.6$). Some
\mbox{metal-rich} clusters of our sample are observed outside the
bulge, i.e. at the distances from the Galactic center larger than
4.5~Kpc: NGC\,104 ([Fe/H]$=\!-1.0$,
[Mg/Fe]$=\!0.6$,~(Sharina et al.,~2018), NGC\,6362 ([Fe/H]$_{\rm
our}=\!-1.45$, [Mg/Fe] = 0.4). Bica et al.~(2016) concluded that the
distributions of the abundances of alpha-process elements (O,
Mg, Si, and Ca) depending on [Fe/H] in the bulge clusters and
field stars are similar.

Abundances of chemical elements provide important information
about supernovae (SNe) and other stars that contributed to the
formation of the chemical composition of the studied objects and
subsystems of the galaxies containing them. Let us compare the
estimated elemental abundances with the model ones in the theory of
chemical evolution by~Kobayashi et al.~(2006). These authors investigated
the production of elements by type II supernovae (SNe\,II) and
hypernovae. Kobayashi et al.~(2006) built chemical evolution models
for the halo, bulge and thick disk of our Galaxy. These three
models show significant differences in chemical abundances (in comparison 
with our corresponding errors
of their determination) for O, Mg, Ca, Ti, Cr and Mn only in the
me\-tal\-li\-ci\-ty range ${\rm [Fe/H]}\!\le\!-2.5$ and \mbox{${\rm
[Fe/H]}\!\ge\!1.0$} (Fig.~32 in~Kobayashi et al., 2006).

Let us consider the distribution of the model ele\-men\-tal abundances
depending on  [Fe/H] presented in Kobayashi et al.~(2006) for the
metallicity range \mbox{$ -2 \le {\rm [Fe/H]} \le -1 $}~dex, where
the objects of our sample are mainly located. The elements  O  and
Mg  were produced mainly in the process of the hydrostatic combustion
in SNe\,II, and their abundances depend on the supernova model. In
the paper by~Kobayashi et al.~(2006)  [O/Fe]$\sim\!0.42$ at
\mbox{${\rm [Fe/H]} \! =\!-1.0$}~dex and is approximately constant in
the range \linebreak \mbox{$ -2\! \le \!{\rm[Fe/H]}\! \le\! -1 $}. Let us
recall that we set the values of \mbox{${\rm [O/Fe]} \sim [0.3;0.5]$}  
for the studied objects. These values are in agreement with the
results of~Kobayashi et al.~(2006).

Model abundances of other $\alpha$-process elements are 
approximately constant in this metallicity range: \mbox{${\rm
[Mg/Fe]}\sim 0.49$}~dex, \mbox{${\rm [Ca/Fe]}\sim
0.27$--$0.39$}~dex (Kobayashi et al., 2006). The observed reduced
abundances [El/Fe] of the alpha-process elements can be explained by 
 SNe\,Ia bursts, which produce mainly iron, and also by the
bursts of \mbox{low-mass} SNe\,II (\mbox{13--15}~$M_{\odot}$)
(Kobayashi et al., 2006). Our results for Mg and Ca 
are consistent with the model data
(Fig.~\ref{fig_Hist}) for the majority of the studied objects. 
We have determined negative [Mg/Fe] values for
NGC\,2419, NGC\,6235, NGC\,6341, NGC\,6779 and NGC\,7078. All
these clusters, except NGC\,6235, have low metallicity ${\rm
[Fe/H]}\le-2$~dex. NGC\,2419 is the most distant massive Galactic
globular cluster with significant chemical anomalies
(see e.g.~Sharina et al.~(2013) and references therein).
A negative [Ca/Fe] value was obtained by us only for NGC\,6544.

The average model [Ti/Fe] abundance is $-0.1$~dex
(Kobayashi et al., 2006). The authors note that the observed values
are systematically higher than the model ones by approximately 0.4~dex.
The [Ti/Fe] values determined by us are basically larger
than zero. Negative values were derived for NGC\,6171, NGC\,6218,
NGC\,6235, NGC\,6266, NGC\,6342, \linebreak NGC\,6652. A part of
these objects reside in the bulge: NGC\,6171, NGC\,6235,
NGC\,6342.

The main sources of carbon and nitrogen are the asymptotic
branch, Wolf--Rayet, and low-mass stars with masses less than the
mass of the Sun. The contribution of these sources is not included
in the calculations by~Kobayashi et al.~(2006). Model carbon abundances
vary approximately from $-0.25$ to $-0.2$~dex 
in the metallicity range
\mbox{$-2\!\le{\rm [Fe/H]}\!\le -1$}~dex. The carbon abundances determined 
in this paper, as well as the observational data for filed stars
from the literature~(Kobayashi et al., 2006) vary in a wide range
(Fig.~\ref{fig_Hist}).

Abundances of the odd \mbox{$Z$-elements} (Na, Al and Cu) depend
strongly on metallicity~(Kobayashi et al., 2006). Note that, for Na,
it is necessary to take into account \mbox{non-LTE} effects in the
atmospheres of stars with ${\rm [Fe/H]}\le-2.0$~dex. Model
abundances of Na increase approximately from $-0.1$ to $0.3$~dex in
the metallicity range $-2 \le{\rm[Fe/H]}\le -1$~(Kobayashi et al., 2006). 
The abundances of Na were confidently determined by us not for
all the sample objects, since the only way to accomplish this task was
to use weak lines Na\,I\,5682\AA, 5688\AA . Furthermore, the
spectra from~Schiavon et al.~(2005) have a relatively low $S/N$ in the
red spectral range. Our estimates are within the range $-0.2
\le{\rm [Na/Fe]}\le 0.7$ (Table~\ref{tab_abund}).

Model abundances of the iron group elements Cr and Mn are
approximately constant in the metallicity range $-2 \le{\rm
[Fe/H]}\le -1$: $-0.5 \le{\rm[Mn/Fe]}\le -0.6$, ${\rm[Cr/Fe]} \sim
0.1$. Spectral lines of Mn and Cr are rather weak in the used spectra.
It can be seen in Fig.~\ref{fig_Hist} that our estimates of the Mn
and Cr abundances vary widely, and the objects with ${\rm
[Mn/Fe]}\ge0$ are NGC\,2298, NGC\,2419, NGC\,5986, NGC\,6522 (Table~\ref{tab_abund}).

Matteucci et al.(2019) have presented chemical evolution models for
the bulge and the inner Galactic disk. The
models by Matteucci et al. differ by the star-formation rate and
efficiency, initial mass function~[Salpeter (1955),
Calamida et al. (2015), Kroupa et al. (1993)]
and by the sources of chemical evolution yields of
SNe\,Ia. As it can be seen in figures presented
by~Matteucci et al.(2019), that the theoretical distribution
[Mg/Fe] versus [Fe/H] for different models, except the model for
the disk, is approximately the following: \mbox{${\rm [Mg/Fe]}\sim0.43$} at
\mbox{${\rm [Fe/H]} =-2.0$}, then the rise up to the values ${\rm
[Mg/Fe]}\sim0.5$ at \mbox{${\rm [Fe/H]} = -1.3$},  and then the
decline to the previous value ${\rm [Mg/Fe]}\sim0.43$ at
\mbox{${\rm [Fe/H]} = -0.7$}.  Our Mg abundances for the sample
clusters in the bulge and for NGC\,104 and NGC\,6362 are generally
consistent  with this model within our errors of the abundances
determination (Figs.~~\ref{fig_abund}, \ref{fig_Elem}).

\section{CONCLUSION AND FURTHER PERSPECTIVES}
\label{discussion} \mbox{Long-slit} \mbox{medium-resolution}
\mbox{integrated-light} spectra of 26 Galactic globular clusters
were used to determine the age, helium mass fraction ($Y$),
metallicity ([Fe/H]), and abundances of chemical elements C, O,
Na, Mg, Ca, Ti, Cr and Mn. Our method, described in
Section~\ref{method} was applied. It was supplemented in this
paper by automatically taking into account the microturbulence
velocity when calculating the spectra of stars in the clusters. 
The method uses medium-resolution integrated-light spectra of clusters
\mbox{($\lambda/ \delta \lambda \ge 1000$)} in a wide spectral
range \mbox{($\lambda \sim [3900; 5000]$~\AA)} and, therefore, can
be applied not only to the study of Galactic, but also bright
extragalactic objects. The choice of the optimal isochrone for
calculating synthetic spectra of clusters is made by matching
the shape and intensity of the observed and theoretical
Balmer line profiles, as well as by reproducing the
observed ratio of the Ca\,I and Ca\,II lines.

In this article we demonstrate the agreement of our results for Galactic
globular clusters with the corresponding literature values obtained
from stellar photometry and spectroscopy of the brightest stars in
the objects with \mbox{high-resolution} spectrographs, as well as
from the studies of integrated spectra of the clusters. The
reasons for the significant deviations of the obtained ages and
metallicities of the clusters from the corresponding literature
values are discussed in Section~4. The systematic difference
between our estimates of [C/Fe] and the literature data obtained
by spectroscopy of the brightest stars in the objects is
interpreted by us as the effect of changes of the chemical
composition in the atmospheres of stars during their evolution.

It was shown that the results of our determination of chemical
abundances at a given metallicity agree with the corresponding
values in the models of che\-mi\-cal evolution by~Kobayashi et al.~(2006)
and~Matteucci et al.(2019).

In the future, we plan to study the influence of
\mbox{non-LTE} effects on the elemental abundances obtained using
\mbox{integrated-light} spectra. Also, we plan to investigate the influence
of the stellar mass function on the integrated spectra of
clusters.

\begin{acknowledgments}
We thank the anonymous referee for the comments, that allowed us to
improve the paper. 
We thank A.~I.~Kolbin  for help in preparing the figures in
Section~3.1 of the article.
\end{acknowledgments}

\section*{FUNDING}
The work was supported by the grant RFBR \linebreak \mbox
{18--02--00167}. Sh.~N.~N. appreciates the grant RFBR
18--42--160003. The work of VVSh was partially funded by the subsidy 
N 0671-2020-0052 to KFU for the scientific activities.

\section*{CONFLICT OF INTEREST}
The authors declare no conflict of interest regarding this paper.

{\it \hspace{3cm}Translated by Sharina M.E.}

\end{document}